\documentclass[aps,pra,reprint,superscriptaddress,twocolumn]{revtex4}
\usepackage[dvipdfmx]{graphicx}
\usepackage{graphicx}
\usepackage{color}
\usepackage[T1]{fontenc}
\usepackage{textcomp}
\usepackage{bm}
\usepackage{amsmath,amssymb,amsthm}
\usepackage{comment}
\usepackage{ulem}
\usepackage{grffile}
\usepackage{algorithm}
\usepackage{algpseudocode}

\newcommand{\cket}[1]{\left|#1\right\rangle}
\newcommand{\bra}[1]{\left\langle#1\right|}
\newcommand{\bracket}[2]{\left\langle#1|#2\right\rangle}

\begin{document}


\title{Persistent-current states originating from the Hilbert space fragmentation in momentum space}

\author{Masaya Kunimi}
\email{kunimi@rs.tus.ac.jp}
\thanks{Present address : Department of Physics, Tokyo University of Science, Shinjuku, Tokyo, 162-8601,  Japan.}
\affiliation{Department of Photo-Molecular Science, Institute for Molecular Science, National Institutes of Natural Sciences, Myodaiji, Okazaki 444-8585, Japan}
\author{Ippei Danshita}
\email{danshita@phys.kindai.ac.jp}
\affiliation{Department of Physics, Kindai University, Higashi-Osaka, Osaka 577-8502, Japan}


\date{\today}

\begin{abstract}
Hilbert space fragmentation (HSF) is a phenomenon wherein the Hilbert space of an isolated quantum system splits into exponentially many disconnected subsectors. The fragmented systems do not thermalize after long-time evolution because the dynamics are restricted to a small subsector. Inspired by recent developments of the HSF, we construct the Hamiltonian that exhibits the HSF in momentum space. We show that persistent-current (PC) states emerge due to the HSF in the momentum space. We also investigate the stability of the PC states against the random potential, which breaks the structure of the HSF, and find that the decay rate of the PC is almost independent of the current velocity.
\end{abstract}

\maketitle

\section{Introduction}\label{sec:Introduction}
Persistent current (PC) states, which have an infinitely long lifetime, are one of the most counter-intuitive phenomena in quantum many-body physics. These phenomena have been studied in various contexts, such as superfluid helium~\cite{Langer_Reppy1970,Reppy1992}, superconductors~\cite{Deaver1961,Doll1961}, mesoscopic physics~\cite{Buttiker1983,Chandrasekhar1991,Mailly1993}, and superfluids of ultracold bosonic and fermionic gases~\cite{Ryu2007,Ramanathan2011,Moulder2012,Beattie2013,Kumar2017,Cai2022,Pace2022,Amico2021,Amico2022}. The PC states are discussed in the ground state for the mesosropic systems \cite{Buttiker1983,Chandrasekhar1991,Mailly1993} and superfluid {}$^3$He \cite{Volovik2003book,Leggett2006}, and in the excited states for superconductivity and superfluidity.
In the latter case, the macroscopic currents flow persistently without any external drive \cite{Bohm1949,Leggett2006}. In such a situation, the PC is a kind of nonequilibrium phenomenon.

\begin{figure}[t]
\centering
\includegraphics[width=8.5cm,clip]{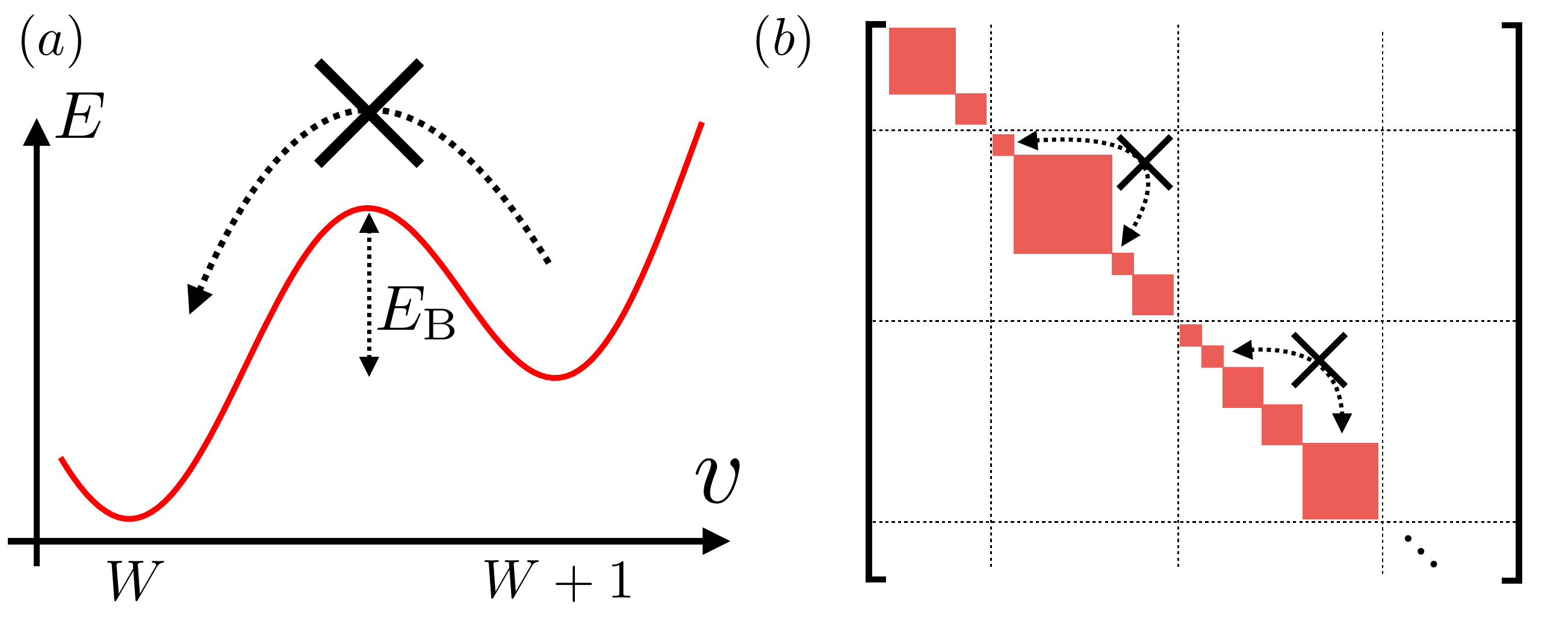}
\caption{(a) Schematic of the typical energy landscape of superfluids as a function of the current  $v$. $E_{\rm B}$ is the energy barrier and $W$ is the winding number. The $W+1$ winding number state is metastable due to a large energy barrier. (b) Schematic of the matrix representation of the fragmented Hamiltonian. The square regions surrounded by dotted lines represent the symmetry sectors, and the red blocks represent disconnected subsectors. The transition between the disconnected subsectors is prohibited.}
\label{fig:hsf_schematic}
\vspace{-0.75em}
\end{figure}

In superconductors and superfluids, the emergence of PCs can be attributed to the existence of a large energy barrier, which prevents the supercurrent from decaying. In this case, the lifetime of the supercurrent is much longer than the experimental timescales [see Fig.~\ref{fig:hsf_schematic}(a)]. However, the situation changes when quantum and/or thermal fluctuations are significant in low-dimensional systems. In these situations, the supercurrent decays due to macroscopic quantum tunneling \cite{Kagan2000,Buchler2001,Polkovnikov2005,Rey2007,Danshita2010,Danshita2012,Danshita2013} and/or thermally activated phase slips \cite{LangerFisher1967,LangerAmbegaokar1967,McCumber1970}. The decay rate depends on the current velocity and system temperature \cite{Kagan2000,Buchler2001,Polkovnikov2005,Danshita2012}. 

The issue of thermalization in isolated quantum many-body systems has become a realistic problem thanks to the developments of various quantum simulators, such as cold atoms or molecules in optical lattices \cite{Bloch2008,Gross2017,Schafer2020}, Rydberg atoms in optical tweezers \cite{Bernien2017,Browaeys2020}, trapped ions \cite{Blatt2012,Monroe2021}, and superconducting qubits \cite{Wendin2017,Kjaergaard2020,Altman2022}. The eigenstate thermalization hypothesis (ETH) is an essential concept of thermalization in isolated systems \cite{Deutsch1991,Srednicki1994,Rigol2008}. If the strong version of the ETH is satisfied, i.e., all eigenstates are thermal, the system thermalizes after long-time unitary evolution starting with any initial conditions \cite{DAlessio2016,Mori2018}. It is also known that some systems do not satisfy the strong ETH. Such types of systems are referred to as nonergodic systems. Typical examples are quantum integrable systems \cite{Rigol2007,Rigol2009,Rigol2009_2,Cassidy2011,Vidmar2016} and Anderson or many-body localized systems \cite{Nandkishore2015,Altman2015,Abanin2019}. Recently, quantum many-body scar \cite{Turner2018,Turner2018_2,Papic2021a,Serbyn2021,Moudgalya2022} and Hilbert space fragmentation (HSF) \cite{Tomasi2019_2,Sala2020,Khemani2020,Moudgalya2021,Moudgalya2022,Scherg2021,Kohlert2023} have been found as novel ergodicity-broken systems. The HSF occurs due to nontrivial conserved quantities such as dipole operator \cite{Sala2020,Khemani2020} and domain wall number operator \cite{Yang2020,Yoshinaga2022}. These operators work as kinetic constraints to the systems. Consequently, the Hilbert space splits into an exponential number of small subsectors, leading to the break down of the ergodicity [see Fig.~\ref{fig:hsf_schematic}(b)]. The minimum dimension among the fragmented subsectors is one. A subsector with the minimum dimension is referred to as a frozen subsector or frozen state because the dynamics are completely frozen if we choose the frozen state as an initial state.

In this paper, we show an altenative mechanism of PCs inspired by the recent developments of understanding of nonergodic systems. In the typical PCs, the decay of the current is suppressed due to the large energy barrier [see Fig.~\ref{fig:hsf_schematic}(a)]. On the other hand, in our mechanism, the decay of the current is prohibited by the HSF in the momentum space, which is caused by the kinetic constraints in the momentum space (see Fig.~\ref{fig:hsf_schematic}(b) and Ref.~\cite{Mukherjee2021}). This is in stark contrast to the ordinary HSF, in which the particles or spin degrees of freedom are localized in the real space so that PC states are absent. Using the exact diagonalization (ED) method, we numerically show that the HSF indeed occurs in the momentum space, and a PC state can be analytically constructed. We investigate the stability of the PC states against the disorder, and find that the decay rate of the PC states hardly depends on the current velocity, making a clear contrast with the behavior of the conventional PCs.

This paper is organized as follows: In Sec.~\ref{sec:model}, we explain how to construct the Hamiltonian that exhibits the HSF in the momentum space. In Secs. ~\ref{subsec:soft-core_boson} and \ref{subsec:spinless_fermions}, we show the ED results for the soft-core bosons and spinless fermions, respectively. In Sec.~\ref{subsec:Persistent_current_states}, we show that the PC states due to the HSF exist in our model. In Sec.~\ref{subsec:stability_Persistent_current_states}, we investigate the stability of the PC state against the random potential. In Sec.~\ref{sec:Summary}, we summarize our results.

\section{Model}\label{sec:model}
We consider soft-core bosons or spinless fermions on a one-dimensional periodic chain. An annihilation (creation) operator at site $j$ is defined by $\hat{a}_j (\hat{a}_j^{\dagger})$, where $j=1,2,\ldots, M$ and $M$ is the number of lattice sites. From the periodic boundary condition, we can introduce the annihilation operator in the momentum space as $\hat{b}_l\equiv (1/\sqrt{M})\sum_{j=1}^Me^{-2\pi i l j/M}\hat{a}_j$, where $l=0,1,\ldots, M-1$ is a crystal momentum. For simplicity, we consider an even-$M$ case only. 

Here, we construct a model that exhibits the HSF in the momentum space. To do this, recall that if the dipole operator like $\sum_j j\hat{a}^{\dagger}_j\hat{a}_j$ commutes with the Hamiltonian, the HSF occurs in the real space \cite{Sala2020,Khemani2020,Moudgalya2021}. We may think we can expect that the Hamiltonian commutes with an operator $\hat{K}\equiv \sum_ll\hat{b}^{\dagger}_l\hat{b}_l$, the HSF occurs in the momentum space. This might be true, but we want to avoid this situation because $\hat{K}$ breaks space-inversion and time-reversal symmetries. Instead of considering the operator $\hat{K}$, we consider the operator 
\begin{align}
\hat{Q}\equiv \sum_{l=0}^{M-1}\left(l-\frac{M}{2}\right)^2\hat{b}^{\dagger}_l\hat{b}_l,\label{eq:definition_of_Q}
\end{align}
which preserves the space-inversion and time-reversal symmetries. 

First, we consider the kinetic term of the Hamiltonian. It is easily shown that the standard nearest-neighbor hopping Hamiltonian commutes with $\hat{Q}$:
\begin{align}
\hat{H}_0&\equiv -J\sum_{j=1}^M(\hat{a}^{\dagger}_{j+1}\hat{a}_j+\hat{a}^{\dagger}_j\hat{a}_{j+1})\notag \\
&=-2J\sum_{l=0}^{M-1}\cos(2\pi l/M)\hat{b}^{\dagger}_l\hat{b}_l.\label{eq:hopping_hamiltonian}
\end{align}
Therefore, we adopt $\hat{H}_0$ as a kinetic term of the system. Next, we consider a two-body interaction term $\hat{H}_{\rm int}$. Let us assume that $\hat{H}_{\rm int}$ has a term $\hat{b}^{\dagger}_{l_1}\hat{b}^{\dagger}_{l_2}\hat{b}_{l_4}\hat{b}_{l_3}$. To satisfy the commutation relation $[\hat{Q}, \hat{H}_{\rm int}]=0$, $l_{1,2,3,4}$ must satisfy the condition $(l_1-M/2)^2+(l_2-M/2)^2=(l_3-M/2)^2+(l_4-M/2)^2$. In addition to this requirement, we assume that the interaction Hamiltonian has the space inversion symmetry, i.e., the relation $[\hat{\mathcal{I}}, \hat{H}_{\rm int}]=0$ is satisfied. Here $\hat{\mathcal{I}}$ is the space-inversion operator, which is defined by $\hat{\mathcal{I}}\hat{a}_j\hat{\mathcal{I}}^{-1}=\hat{a}_{M-j}$ for real space and $\hat{\mathcal{I}}\hat{b}_l\hat{\mathcal{I}}^{-1}=\hat{b}_{M-l}$ for momentum space. From the above considerations, we adopt the following two-body interaction Hamiltonian that commutes with $\hat{Q}$ and $\hat{\mathcal{I}}$
\begin{align}
\hat{H}_{\rm int}&=\frac{V}{M}\sum_{s=1}^M(\hat{b}^{\dagger}_{M/2+s}\hat{b}^{\dagger}_{M/2-s-1}\hat{b}_{M/2+s+1}\hat{b}_{M/2-s}+{\rm H.c.}),\label{eq:two-body_interaction_Hamiltonian}
\end{align}
where $V$ represents the interaction strength and ${\rm H.c.}$ denotes the hermitian conjugate. The prefactor $1/M$ is introduced because the total energy should have extensiveness. The total Hamiltonian is given by $\hat{H}\equiv \hat{H}_0+\hat{H}_{\rm int}$. We emphasize that the interaction Hamiltonian (\ref{eq:two-body_interaction_Hamiltonian}) is not a unique choice for satisfying the relation $[\hat{Q}, \hat{H}_{\rm int}]=0$.

\begin{figure*}[t]
\centering
\includegraphics[width=12cm,clip]{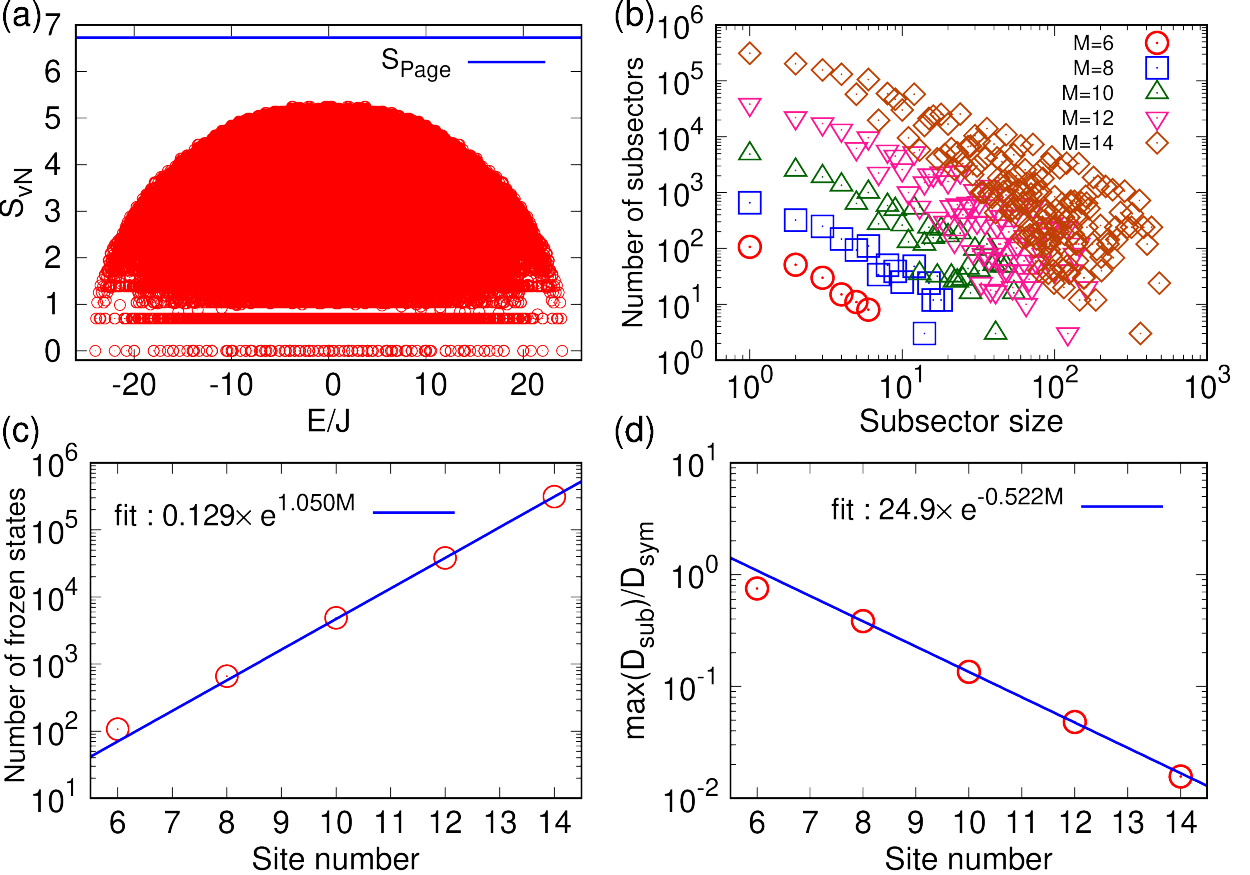}
\caption{ED results for the soft-core bosons. (a) von Neumann entanglement entropy as a function of the eigenenergy for $M=N=12$ and $V=0.1J$. The dimension of $\mathcal{H}_N$ is 1352078. The blue solid line represents the Page value of the maximum symmetry sector $(N,Q,N_{\rm even},\mathcal{I})=(12, 134, 6, +1)$ whose dimension is 3477. (b) Distribution of the subsector size for various system sizes. (c) Site number dependence of the number of frozen states. The blue solid line represents the fit to an exponential function. (d) Site number dependence of the ratio ${\rm max}(D_{\rm sub})/D_{\rm sym}$. The blue solid line represents the fit to an exponential function.}
\label{fig:hsf_figure}
\vspace{-0.75em}
\end{figure*}%

Now, we discuss the conserved quantities of the system. By construction, $\hat{Q}$ and $\hat{\mathcal{I}}$ are conserved. Moreover, since there is the ${\rm U}(1)$ symmetry with respect to the global rotation of the phase of the field operator, the total particle number $\hat{N}\equiv \sum_{j=1}^M\hat{a}^{\dagger}_j\hat{a}_j=\sum_{l=0}^{M-1}\hat{b}^{\dagger}_l\hat{b}_l$ is conserved. We define the Hilbert subspace with fixed particle number; $\mathcal{H}_N\equiv \{\cket{\phi} | \hat{N}\cket{\phi}=N\cket{\phi}\}$, where $N$ is the total particle number of the system. We adopt $\cket{\bm{n}}\equiv \cket{n_0,n_1,\ldots, n_{M-1}}$ as a basis set in $\mathcal{H}_N$, where $n_{l}=0,1,2,\ldots, N_{\rm max}$ is an occupation number of the crystal momentum $l$, and $N_{\rm max}$ is the maximum occupation number. In the case of bosons $N_{\rm max}=N$ whereas in the spinless fermion case $N_{\rm max}=1$. In addition to these conserved quantities, $\hat{N}_{\rm even}\equiv \sum_{l=0,2,\ldots}\hat{b}^{\dagger}_l\hat{b}_l$ is a conserved quantity. Given these four conserved quantities, the symmetry sector is defined by $\mathcal{H}_{N,Q,N_{\rm even},\mathcal{I}}\equiv \{\cket{\phi} | \hat{N}\cket{\phi}=N\cket{\phi}, \hat{Q}\cket{\phi}=Q\cket{\phi}, \hat{N}_{\rm even}\cket{\phi}=N_{\rm even}\cket{\phi}, \hat{\mathcal{I}}\cket{\phi}=\mathcal{I}\cket{\phi}\}$, where $Q$, $N_{\rm even}$, and $\mathcal{I}=\pm1$ are an eigenvalue of $\hat{Q}$, $\hat{N}_{\rm even}$, and $\hat{\mathcal{I}}$, respectively. In this paper, we analyze the above system using the ED method \cite{Sandvik2010,Jung2020}. To perform the ED calculations, we use a basis set incorporating the symmetries. We introduce a basis $\cket{\bar{\bm{n}}; N, Q, N_{\rm even}, \mathcal{I}}\equiv[\sqrt{q(\bar{\bm{n}})}/2](\cket{\bar{\bm{n}}}+\mathcal{I}\hat{\mathcal{I}}\cket{\bar{\bm{n}}})$, where $\cket{\bar{\bm{n}}}$ is the representative state \cite{Sandvik2010,Jung2020} and $\cket{\bar{\bm{n}}}$ satisfies $\hat{N}\cket{\bar{\bm{n}}}=N\cket{\bar{\bm{n}}}, \hat{Q}\cket{\bar{\bm{n}}}=Q\cket{\bar{\bm{n}}}$, and $\hat{N}_{\rm even}\cket{\bar{\bm{n}}}=N_{\rm even}\cket{\bar{\bm{n}}}$. The definition of $q(\bar{\bm{n}})$ depends on the particle statistics. In the soft-core boson case, $q(\bar{\bm{n}})$ is $1$ for $\hat{\mathcal{I}}\cket{\bar{\bm{n}}}=\cket{\bar{\bm{n}}}$ or 2 for $\hat{\mathcal{I}}\cket{\bar{\bm{n}}}\not=\cket{\bar{\bm{n}}}$. In the spinless fermion case, owing to the Fermionic sign, we obtain
\begin{align}
\hat{\mathcal{I}}\cket{\bm{n}}&=\hat{\mathcal{I}}\cket{n_0,n_1,\ldots,n_{M-1}}\notag \\
&=(-1)^{S(\bm{n})}\cket{n_0,n_{M-1},n_{M-2},\ldots, n_2,n_1},\label{eq:space_inversion_for_Fermion}\\
S(\bm{n})&\equiv \frac{1}{2}(N-n_0)(N-n_0-1).\label{eq:definition_of_Fermion_sign_space_inversion}
\end{align}
From this result, when $q(\bar{\bm{n}})=1$, we have $\hat{\mathcal{I}}\cket{\bar{\bm{n}}}=\pm\cket{\bar{\bm{n}}}$ in contrast to the bosonic case. For example, $\hat{\mathcal{I}}\cket{0,1,0,1}=-\cket{0,1,0,1}$ and $\hat{\mathcal{I}}\cket{1,0,1,0}=\cket{1,0,1,0}$ hold. Therefore, we define $q(\bar{\bm{n}})=1$ for $\hat{\mathcal{I}}\cket{\bar{\bm{n}}}=\pm\cket{\bar{\bm{n}}}$ and $q(\bar{\bm{n}})=2$ for $\hat{\mathcal{I}}\cket{\bar{\bm{n}}}\not\propto\cket{\bar{\bm{n}}}$.

\section{Results}\label{sec:Results}
\subsection{Soft-core bosons}\label{subsec:soft-core_boson}

Here, we verify that the Hamiltonian $\hat{H}=\hat{H}_0+\hat{H}_{\rm int}$ for the soft-core bosons exhibits the HSF in the momentum space. Figure~\ref{fig:hsf_figure}(a) shows the bipartite entanglement entropy (EE) in the momentum space for $M=N=12$ \cite{Lundgren2014,Schindler2022}. The subsystem $A(B)$ is defined by $l=0,1,\ldots, M/2-1 \;(M/2,M/2+1,\ldots, M-1)$. We can see a broad distribution of the EE. There are many low-entangled eigenstates even in the center of the spectra. This behavior is in contrast to the ergodic systems. In addition to this property, the maximum value of the EE is significantly lower than the Page value \cite{Page1993}. This is a signature of ergodicity breaking. To verify the HSF that occurs in this system quantitatively, we investigate the distribution of the subsectors. Figure~\ref{fig:hsf_figure}(b) shows the distribution of the subsector size for various system sizes. We find a broad distribution of the subsectors. Here, we focus on the number of the frozen states, whose subsector dimension is one. Figure~\ref{fig:hsf_figure}(c) shows the system size dependence of the number of frozen states. We find that the number of frozen states scales as an exponential function of the system size. In Appendix \ref{sec:analytical_estimation_frozen_state}, we analytically estimate the number of frozen states. We also plot the ratio ${\rm max}(D_{\rm sub})/D_{\rm sym}$ in Fig.~\ref{fig:hsf_figure}(d), where ${\rm max}(D_{\rm sub})$ is the dimension of the maximum subsector within the full Hilbert space and $D_{\rm sym}$ is the dimension of the symmetry sector that the maximum subsector belongs to. The ratio ${\rm max}(D_{\rm sub})/D_{\rm sym}$ decreases as an exponential function of the system size. This behavior is consistent with the strong version of the HSF \cite{Sala2020}. From these results, we conclude that the Hamiltonian (\ref{eq:two-body_interaction_Hamiltonian}) exhibits the HSF in the momentum space.

\begin{figure*}[t]
\centering
\includegraphics[width=12cm,clip]{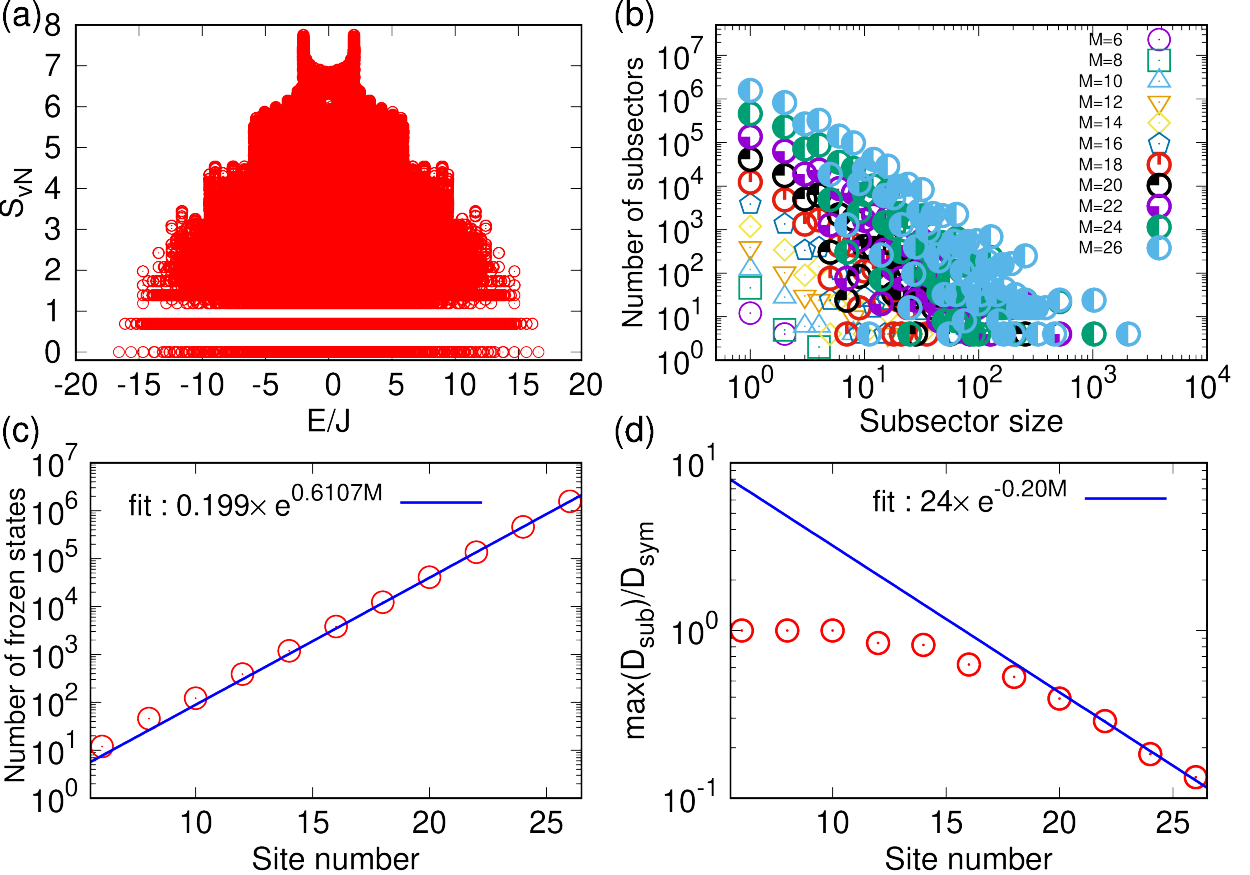}
\caption{ED results for the spinless fermions. (a) von Neumann entanglement entropy as a function of the eigenenergy for $M=26$ and $N=13$, and $V=0.1J$. The dimension of $\mathcal{H}_N$ is 10400600. (b) Distribution of the subsector size for various system sizes. The particle number is fixed to $N=M/2$. (c) Site number dependence of the number of frozen states. The blue solid line represents the fit to an exponential function. (d) Site number dependence of the ratio ${\rm max}(D_{\rm sub})/D_{\rm sym}$. The blue solid line represents the fit to an exponential function.}
\label{fig:hsf_fermion}
\vspace{-0.75em}
\end{figure*}%

\subsection{Spinless fermions}\label{subsec:spinless_fermions}

Here, we show the exact diagonalization results for the spinless fermion case. The Hamiltonian is the same as Sec.~\ref{subsec:soft-core_boson}, except that the operators are spinless fermions. Figure~\ref{fig:hsf_fermion} shows the numerical results of the von Neumann entanglement entropy, distribution of the subsector size, the number of frozen states, and ${\rm max}(D_{\rm sub})/D_{\rm sym}$ for $V=0.1J$, $M=26$, and $N=13$. We find that all results are qualitatively the same as the case of the soft-core bosons. Therefore, we conclude that the HSF in the momentum space also occurs in the case of the spinless fermion.

\subsection{Persistent current states}\label{subsec:Persistent_current_states}

Here, we show that the system has PC states, which have macroscopic current and infinite lifetime. In the following, we consider the soft-core bosons only. Before discussing the PC states, we define the expression of the current operator $\hat{v}$: 
\begin{align}
\hat{v}&\equiv \hat{v}_{\rm kin}+\hat{v}_{\rm int},\label{eq:definition_of_current_operator_total}\\
\hat{v}_{\rm kin}&\equiv \frac{2dJ}{\hbar}\sum_{l=0}^{M-1}\sin\left(\frac{2\pi l}{M}\right)\hat{b}^{\dagger}_l\hat{b}_l,\label{eq:definition_of_current_operator_hopping_part}\\
\hat{v}_{\rm int}&\equiv -\frac{idV}{M\hbar}\sum_{s=1}^M\sum_{l=0}^{M-1}(f_{-l}\hat{c}^{\dagger}_{s+l}\hat{c}^{\dagger}_{-s-1+l}\hat{c}_{s+1+l}\hat{c}_{-s-l}-{\rm H.c.}),\label{eq:definition_of_current_operator_interaction_part}\\
f_l&\equiv 
\begin{cases}
(-1)^{l}\left[-\dfrac{1}{2}-\dfrac{i}{2}\cot(\pi l/M)\right],\quad l\not=0~{\rm mod }~M,\\
-\dfrac{1}{2},\quad l=0~{\rm mod}~ M,
\end{cases}
\label{eq:definition_of_f_l}
\end{align}
where $d$ is the lattice spacing, $\hat{c}_l\equiv \hat{b}_{M/2+l}$. Since our Hamiltonian has nontrivial interaction terms, the current operator from the interaction part $\hat{v}_{\rm int}$ exists unlike the typical situations. Here, we derive the expression of the current operator. To do this, we introduce the Hamiltonian with a uniform phase twist:
\begin{align}
\hat{H}(\theta)&\equiv \hat{U}^{\dagger}(\theta)\hat{H}\hat{U}(\theta),\label{eq:definition_phase_twisted_Hamiltonian}\\
\hat{U}(\theta)&\equiv e^{-i\theta\hat{X}},\label{eq:definition_of_phase_twist_operator}\\
\hat{X}&\equiv \sum_{j=1}^Mj\hat{a}^{\dagger}_j\hat{a}_j,\label{eq:definition_of_dipole_operator_real_basis}
\end{align}
where $\theta$ is a real number. From Eqs.~(\ref{eq:definition_of_phase_twist_operator}) and (\ref{eq:definition_of_dipole_operator_real_basis}), the following relation holds: $\hat{U}^{\dagger}(\theta)\hat{a}_j\hat{U}(\theta)=e^{-i\theta j}\hat{a}_j$. The current operator is defined by
\begin{align}
\hat{v}&\equiv -\frac{d}{\hbar}\left.\frac{\partial \hat{H}(\theta)}{\partial\theta}\right|_{\theta=0}.\label{eq:definition_of_the_current_operator}
\end{align}
Using this definition, we obtain the expressions of the current operator (\ref{eq:definition_of_current_operator_total}), (\ref{eq:definition_of_current_operator_hopping_part}), and (\ref{eq:definition_of_current_operator_interaction_part}).

Now, we focus on the frozen states. Because the subsector dimension is one in the frozen subsector, the state $\cket{\bar{\bm{n}};N,Q,N_{\rm even},\mathcal{I}}$ is an eigenstate of the Hamiltonian. There are two kinds of frozen states. One is the case $q(\bar{\bm{n}})=1$. In this case, $\cket{\bar{\bm{n}}}$ is an eigenstate of the space inversion operator. From the space inversion symmetry, the current must be zero because the current operator $\hat{v}$ has odd parity under the space inversion, i.e., $\hat{\mathcal{I}}\hat{v}\hat{\mathcal{I}}^{-1}=-\hat{v}$. In the case of $q(\bar{\bm{n}})=2$, the frozen state is a superposition of $\cket{\bar{\bm{n}}}$ and $\hat{\mathcal{I}}\cket{\bar{\bm{n}}}$. Utilizing these facts, we can show that the state $\cket{\bar{\bm{n}}}$ can have a finite current and infinite lifetime. To show this, we use the relations
\begin{align}
\cket{\bar{\bm{n}}}&=\frac{1}{\sqrt{2}}(\cket{\bar{\bm{n}};N,Q,N_{\rm even},+1}+\cket{\bar{\bm{n}};N,Q,N_{\rm even},-1})\notag \\
&\equiv \frac{1}{\sqrt{2}}(\cket{+}+\cket{-}).\label{eq:representative_state_written_by_+_and_-_state}
\end{align}
If we choose the initial condition as $\cket{\psi(0)}=\cket{\bar{\bm{n}}}$, the expectation value of the current operator is given by
\begin{align}
\bra{\psi(t)}\hat{v}\cket{\psi(t)}&=\frac{1}{2}e^{i(E_+-E_-)t/\hbar}\bra{+}\hat{v}\cket{-}+{\rm c.c.},\label{eq:expectation_value_of_current_operator_at_time_t}
\end{align}
where $E_{\pm}$ is the eigenvalue of the states $\cket{\pm}$. This result implies that if $E_+=E_-$ and $\bra{+}\hat{v}\cket{-}\not=0$, the expectation value of the current is finite and time independent. When the expectation value of the current has macroscopic value,  this is equivalent to the PC state. For example, the state $\cket{\bm{n}}=\cket{0,N,0,\ldots,0}$ satisfies the above conditions. The expectation value of the current operator scales as $O(N/M)$, which survives in the thermodynamics limit. We also find that if $E_+\not=E_-$ and $\bra{+}\hat{v}\cket{-}\not=0$, the expectation value of the current persistently oscillates in time with the period $2\pi\hbar/|E_+-E_-|$. In this case, the current expectation value typically scales as $O(1/M)$. These current oscillating states do not carry the macroscopic current. We also note that the number of finite-current states can be estimated analytically. See Appendix~\ref{sec:analytical_estimation_finite-current} for the details.

\subsection{Stability of the persistent current states}\label{subsec:stability_Persistent_current_states}

Finally, we discuss the stability of the PC states against disorder. Here, we consider two types of disorders. One is the diagonal disorder in the momentum space (or disoredered hopping in the real space). The perturbation Hamiltonian is given by 
\begin{align}
\hat{H}'\equiv \sum_lT_l\hat{b}^{\dagger}_l\hat{b}_l\equiv \sum_{j,k}\tilde{T}_{j k}\hat{a}^{\dagger}_j\hat{a}_k,\label{eq:perturbation_diagonal_in_momentum_space}
\end{align}
where $T_l$ is an arbitrary real number and $\tilde{T}_{j k}\equiv (1/M)\sum_lT_le^{2\pi i l(j-k)/M}$. Owing to the properties of the HSF, if we add a disorder potential with only diagonal matrix elements in the momentum space, the structure of the HSF preserves. Therefore, the PC states are stable against the diagonal perturbation in the momentum space. However, the diagonal disorder in the real space (random hopping in the momentum space) breaks the structure of the HSF in the momentum space. We consider the Hamiltonian 
\begin{align}
\hat{H}_{\rm rand}\equiv \sum_jU_j\hat{a}^{\dagger}_j\hat{a}_j=\sum_{l,l'}\tilde{U}_{l-l'}\hat{b}^{\dagger}_{l'}\hat{b}_l,\label{eq:perturbation_diagonal_in_real_space}
\end{align}
where $U_j$ is a real number that obeys the uniform distribution on the interval $[-W/2,W/2] \; (W>0)$ and $\tilde{U}_{l-l'}\equiv (1/M)\sum_j U_j e^{2\pi i(l-l')j/M}$. To investigate the stability of the PC states, we numerically solve the time-dependent Schr\"odinger equation with the Hamiltonian $\hat{H}_{\rm tot}\equiv \hat{H}+\hat{H}_{\rm rand}$ starting with the initial condition $\cket{\psi(0)}=\cket{0,N,0,\ldots, 0}$. We use the Krylov subspace method for calculating the time-evolution operator $e^{-i\hat{H}_{\rm tot}t/\hbar}$ \cite{Hochbruck1996,Paeckel2019}. The following numerical results are averaged over 750 realizations of the random potential.

\begin{figure}[t]
\centering
\includegraphics[width=8.6cm,clip]{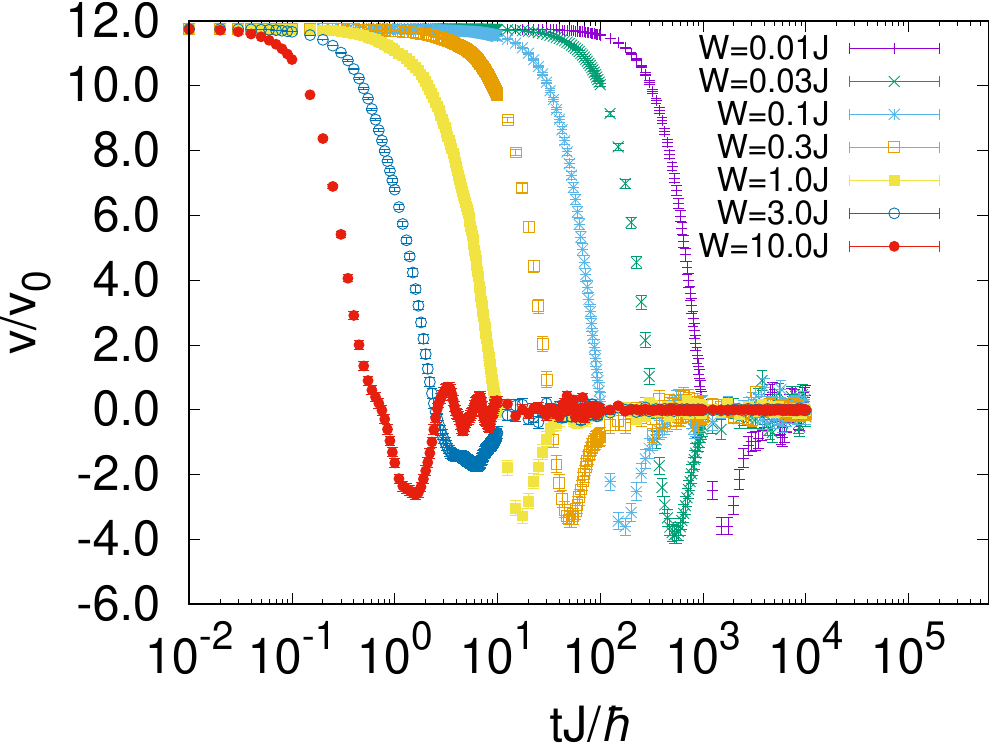}
\caption{Expectation value of the current operator $\hat{v}$ for $M=N=10$ and $V=0.1J$. The initial condition is $\cket{\psi(0)}=\cket{0,10,0,\ldots,0}$. $v_0\equiv dJ/\hbar$. The error bars represent the standard error of the mean.}
\label{fig:current_psi_0_10_v0.1}
\vspace{-0.75em}
\end{figure}%

\begin{figure}[t]
\centering
\includegraphics[width=8.6cm,clip]{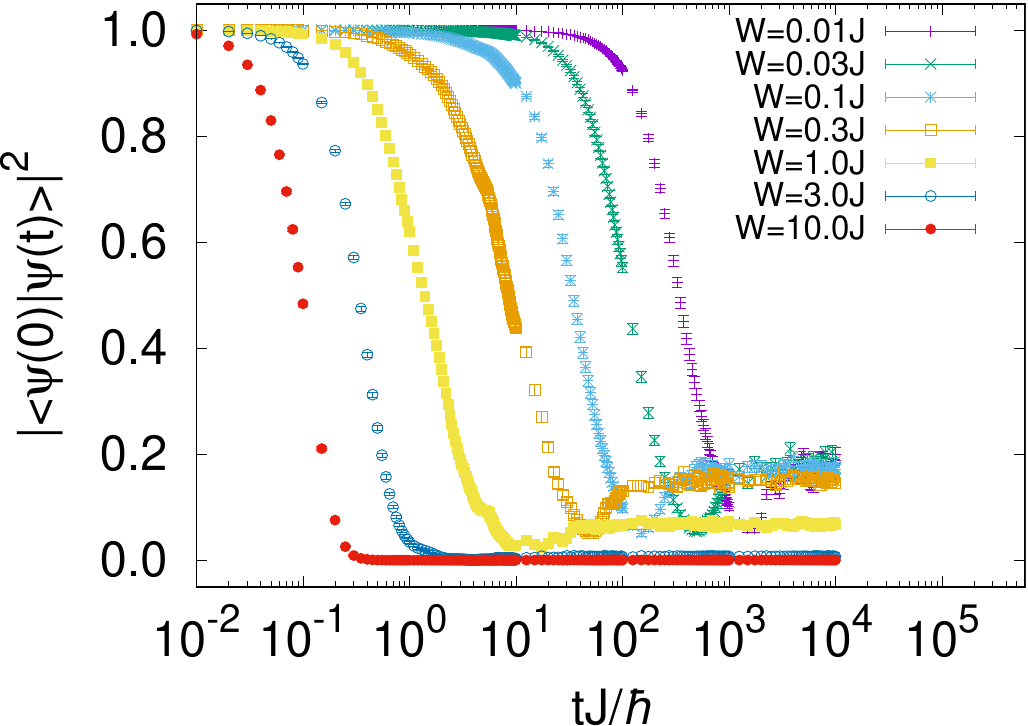}
\caption{Fidelity for $M=N=10$ and $V=0.1J$. The initial condition is $\cket{\psi(0)}=\cket{0,10,0,\ldots, 0}$. The error bars represent the standard error of the mean.}
\label{fig:fidelity_psi_0_10_v0.1}
\vspace{-0.75em}
\end{figure}%

\begin{figure}[t]
\centering
\includegraphics[width=8.6cm,clip]{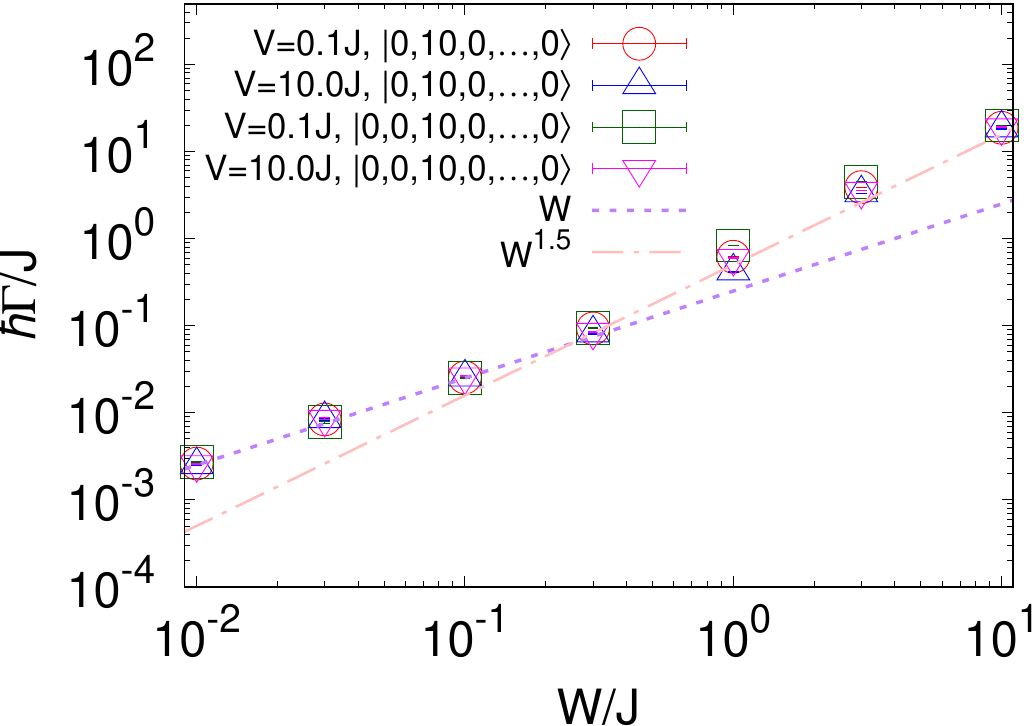}
\caption{Decay rate as a function of the disorder strength for $M=N=10$ and $\cket{\psi(0)}=\cket{0,10,0,\ldots, 0}$ and $\cket{\psi(0)}=\cket{0,0,10,0,\ldots, 0}$. The dotted green and dashed-dotted line represent $W$ and $W^{1.5}$ for a guide to the eye.}
\label{fig:gamma_0_10}
\vspace{-0.75em}
\end{figure}%

Figure~\ref{fig:current_psi_0_10_v0.1} shows the time evolution of the current for various disorder strengths. As discussed above, the PC states decay after long-time evolution. The decay is faster for the larger disorder strength. This property can be seen in the fidelity shown in Fig.~\ref{fig:fidelity_psi_0_10_v0.1}, which is defined by the overlap between the initial state $\cket{\psi(0)}$ and the state $\cket{\psi(t)}$. We can see that the decay of the current and that of the fidelity are qualitatively similar. To characterize the timescale of the decay quantitatively, we perform fitting to the fidelity data $|\langle\psi(0)|\psi(t)\rangle|^2$ with a function $Ae^{-\Gamma t}$, where the amplitude $A$ and decay rate $\Gamma$ are fitting parameters \cite{Danshita2012}. See Appendix~\ref{sec:details_of_fitting} for the details. Figure~\ref{fig:gamma_0_10} shows the decay rate as a function of the disorder strength. We find that the decay rate is proportional to the disorder strength for the small $W$ region. This is consistent with predictions of Fermi's golden rule. According to Fermi's golden rule, the expression of the decay rate is given by $\Gamma\propto \sum_n|\bra{n}\hat{H}_{\rm rand}\cket{\psi(0)}|^2\delta(E_n-E_{\rm ini})$, where $\cket{n}$ and $E_n$ are the eigenstate and eigenenergy of $\hat{H}$, respectively, and $E_{\rm ini}$ is the expectation of the non-perturbative energy of the initial state. In the present case, the matrix elements are proportional to $W^2$, and the delta function part, which can be interpreted as the density of states, is proportional to $1/W$ \cite{Torres-Herrera2014,Tomasi2019}. Therefore, the decay rate is proportional to $W$ in the small $W$ region. In the large $W$ region, we can see a clear deviation from the perturbative results. Although we plot $W^{1.5}$ for a guide to the eye, we do not identify the origin of this behavior. 

In the end of this section, we discuss the difference between the conventional PC in superfluids and our PC states. There are two main differences. One is its mechanism. In the superfluids, the PC states are protected by the existence of a large energy barrier as shown in Fig.~\ref{fig:hsf_schematic} (a). This is due to the mean-field nature. On the other hand, the origin of our PC is the kinetic constraints arising from the nontrivial conserved quantity $\hat{Q}$. This is a purely quantum effect. The other difference is the behavior of the decay rate. For example, in one-dimensional Bose gases, the decay rate depends on the initial state (current) and/or temperature \cite{Kagan2000,Buchler2001,Fertig2005,Danshita2012,Danshita2013,Tanzi2016,Kunimi2017}. However, our PC states show distinct properties. Figure~\ref{fig:gamma_0_10} shows the results for the decay rate as a function of the disorder strength for different initial conditions $\cket{\psi(0)}=\cket{0,N,0,\ldots, 0}$ and $\cket{\psi(0)}=\cket{0,0,N,0,\ldots, 0}$. This result shows that the decay rate of our PC is insensitive to the initial condition. From these facts, our PC states differ from those originating from superfluidity.

\section{Summary}\label{sec:Summary}

In this paper, we proposed a mechanism for the PC states originating from the HSF in the momentum space. The mechanism is different from that of conventional superconductors or superfluids. The decay of the current is prohibited by the kinetic constraint in momentum space arising from the nontrivial conserved quantity $\hat{Q}$. From the exact numerical calculations, we showed that the stability of the PC states against the disorder is different from that of the conventional ones.

Finally, we remark on the experimental feasibility of the model that exhibits the HSF in the momentum space. As shown in this paper, our model is interesting as a theoretical model that exhibits the persistent current originating from the HSF in the momentum space. Unfortunately, it is not easy to realize in the current experimental techniques because we need to engineer the interaction Hamiltonian that conserves the operator $\hat{Q}$. We expect that there is a possibility for the experimental realization of our model in the future based on the trend of progress in quantum simulators.

\begin{acknowledgments}
We thank H. Katsura and Y. Suzuki for useful comments. This work was supported by MEXT Quantum Leap Flagship Program (MEXT Q-LEAP) Grant No. JPMXS0118069021 (M.K. and I.D.), JST FOREST Grant No. JPMJFR202T (I.D.), and JSPS KAKENHI Grants No. JP20K14389 (M.K.), No. JP22H05268 (M.K.), No. JP21H01014 (I.D.), and No. JP18H05228 (I.D.).

\end{acknowledgments}
\appendix
\begin{widetext}

\section{GENERALIZATION OF THE HAMILTONIAN}\label{sec:generralization_Hamiltonian}
In the main text, we discuss the HSF in the momentum space for the specific Hamiltonian. The Hamiltonian is constructed to satisfy the commutation relations
\begin{align}
[\hat{N}, \hat{H}]=0,\; [\hat{Q}, \hat{H}]=0,\; [\hat{N}_{\rm even}, \hat{H}]=0,\; [\hat{\mathcal{I}}, \hat{H}]=0.\label{eq:commutation_relation_for_HSF}
\end{align}
Here, we show that a more general form of the Hamiltonian also satisfies the commutation relations (\ref{eq:commutation_relation_for_HSF}) and exhibits the HSF in the momentum space. 

We can check that the following Hamiltonians satisfies the commutation relations (\ref{eq:commutation_relation_for_HSF}):
\begin{align}
\hat{H}^{(i)}&\equiv \hat{H}_0'+\hat{H}_{\rm int}^{(i)},\;(i=1,2,3),\label{eq:general_form_of_Hamiltonian}\\
\hat{H}_0'&\equiv \sum_{l=0}^{M-1}\epsilon_l\hat{n}_l,\label{eq:single_particle_term_general_Hamiltonian}\\
\hat{H}^{(1)}_{\rm int}&\equiv\sum_{r=1}^{M/2}\frac{V_r^{(1)}}{M}\sum_{s=1}^M(\hat{b}^{\dagger}_{M/2+s}\hat{b}^{\dagger}_{M/2-s-r}\hat{b}_{M/2+s+r}\hat{b}_{M/2-s}+\hat{b}^{\dagger}_{M/2-s}\hat{b}^{\dagger}_{M/2+s+r}\hat{b}_{M/2-s-r}\hat{b}_{M/2+s}),\label{eq:general_Hamiltonian1}\\
\hat{H}_{\rm int}^{(2)}&\equiv\sum_{r=1}^{M/2}\frac{V_r^{(2)}}{M}\sum_{s=1}^M(\hat{b}^{\dagger}_{M/2+s}\hat{b}^{\dagger}_{M/2+s+r}\hat{b}_{M/2-s-r}\hat{b}_{M/2-s}+\hat{b}^{\dagger}_{M/2-s}\hat{b}^{\dagger}_{M/2-s-r}\hat{b}_{M/2+s+r}\hat{b}_{M/2+s}),\label{eq:general_Hamiltonian2}\\
\hat{H}_{\rm int}^{(3)}&\equiv\sum_{r=1}^{M/2}\frac{V_r^{(3)}}{M}\sum_{s=1}^M(\hat{b}^{\dagger}_{M/2+s}\hat{b}_{M/2+s+r}^{\dagger}\hat{b}_{M/2+s+r}\hat{b}_{M/2-s}+\hat{b}^{\dagger}_{M/2-s}\hat{b}_{M/2+s+r}^{\dagger}\hat{b}_{M/2+s+r}\hat{b}_{M/2+s}),\label{eq:general_Hamiltonian3}
\end{align}
where $\hat{b}_l$ and $\hat{b}^{\dagger}_l$ are the annihilation and creation operators of the soft-core boson or spinless fermion, $\epsilon_l=\epsilon_{M-l}\in\mathbb{R}$ is the single-particle dispersion, and $V_r^{(1,2,3)}\in\mathbb{R}$ represents the interaction strength. If we choose $\epsilon_l=-2J\cos(2\pi l/M)$ and $V_r^{(1)}=V\delta_{r,1}$, the Hamiltonian (\ref{eq:general_Hamiltonian1}) reduces to that used in the main text. 

For reference, we show the real-space representation of the Hamiltonian $\hat{H}^{(i)}$. The results are as follows:
\begin{align}
\hat{H}_0'&=\sum_{j,k}\tilde{\epsilon}_{j,k}\hat{a}^{\dagger}_j\hat{a}_k,\label{eq:real_space_representation_one-body_term}\\
\tilde{\epsilon}_{j,k}&\equiv \frac{1}{M}\sum_{l=0}^{M-1}\epsilon_le^{2\pi i l(j-k)/M},\label{eq:definition_of_tilde_epsilon_j_k}\\
\hat{H}_{\rm int}^{(1)}&=\sum_{r=1}^{M/2}\frac{2V_r^{(1)}}{M^2}\sum_{j_1,j_2,j_3,j_4}\delta_{j_1-j_2+j_3-j_4,0\;{\rm mod}\;M }(-1)^{j_1+j_2-j_3-j_3}\cos\left[\frac{2\pi(j_2+j_4)r}{M}\right]\hat{a}^{\dagger}_{j_1}\hat{a}^{\dagger}_{j_2}\hat{a}_{j_4}\hat{a}_{j_3},\label{eq:Hamiltonian1_real_space_representation}\\
\hat{H}_{\rm int}^{(2)}&=\sum_{r=1}^{M/2}\frac{2V_r^{(2)}}{M^2}\sum_{j_1,j_2,j_3,j_4}\delta_{j_1+j_2+j_3+j_4,0\;{\rm mod}\; M}(-1)^{j_1+j_2-j_3-j_4}\cos\left[\frac{2\pi(j_2+j_3)r}{M}\right]\hat{a}^{\dagger}_{j_1}\hat{a}^{\dagger}_{j_2}\hat{a}_{j_4}\hat{a}_{j_3},\label{eq:Hamiltonian2_real_space_representation}\\
\hat{H}_{\rm int}^{(3)}&=\sum_{r=1}^{M/2}\frac{2V_r^{(3)}}{M^2}\sum_{j_1,j_2,j_3,j_4}\delta_{j_1+j_2+j_3-j_4,0\;{\rm mod}\;M}(-1)^{j_1+j_2-j_3-j_4}\cos\left[\frac{2\pi(j_2-j_4)r}{M}\right]\hat{a}^{\dagger}_{j_1}\hat{a}^{\dagger}_{j_2}\hat{a}_{j_4}\hat{a}_{j_3},\label{eq:Hamiltonian3_real_space_representation}
\end{align}
where $\delta_{x,0\;{\rm mod}\;M}=1$ if $x=0\;{\rm mod}\;M$ otherwise 0. From the above expressions, we can easily see that the system has long-range interactions and breaks the translational symmetry.

As an example, we consider the Hamiltonian (\ref{eq:general_Hamiltonian1}) with $\epsilon_l=-2J\cos(2\pi l/M)$ and $V_r^{(1)}=V \;(r\le r_{\rm max}) \text{ and }\text{ otherwise}$ 0 for the soft-core bosons. When $r_{\rm max}=1$, this Hamiltonian coincides with that in the main text. Figure~\ref{fig:hsf_figure_h1_r6} shows the numerical results of the von Neumann entanglement entropy, distribution of the subsector size, the number of frozen states, and ${\rm max}(D_{\rm sub})/D_{\rm sym}$ for $V=0.1J$ and $r_{\rm max}=6$. We can find that the qualitative behavior is almost the same as that shown in Fig.~2 of the main text. This indicates that the other form of the Hamiltonian also exhibits the HSF in the momentum space. Figure~\ref{fig:rmax_dependence_hsf_figure_h1_r7} shows the system size dependence of the number of the frozen states for various $r_{\rm max}$. These results suggest that the number of frozen states decreases for increasing $r_{\rm max}$. The reason for this behavior is that the number of the off-diagonal elements of the Hamiltonian is an increase in function of $r_{\rm max}$. Although the number of the frozen states becomes small for large $r_{\rm max}$, the system size dependence is still exponential. This means that the strong version of the HSF still survives, even in the case of a finite interaction range. 

\begin{figure}[t]
\centering
\includegraphics[width=12cm,clip]{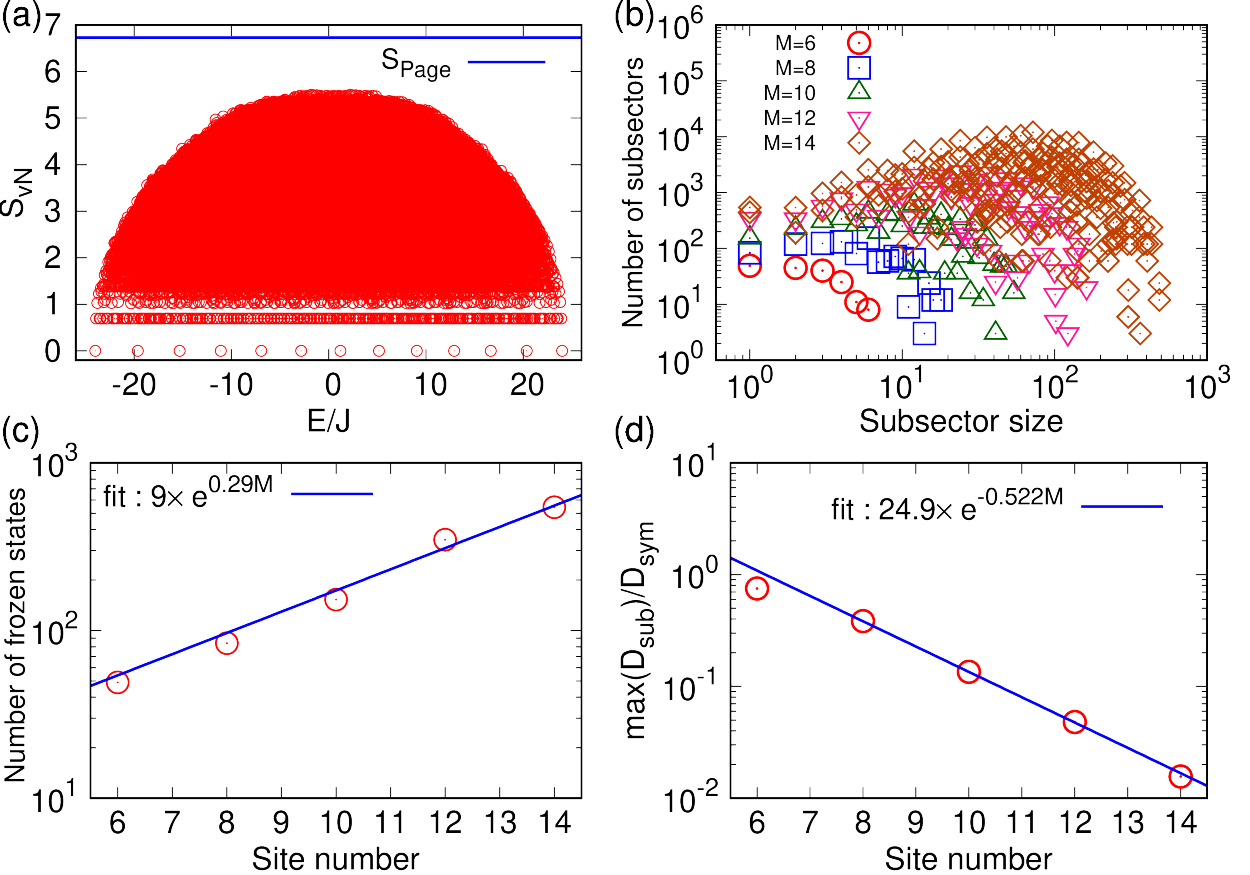}
\caption{ED results for the soft-core bosons. (a) von Neumann entanglement entropy as a function of the eigenenergy for $M=N=12$, $r_{\rm max}=6$, and $V=0.1J$. The dimension of $\mathcal{H}_N$ is 1352078. The blue solid line represents the Page value of the maximum symmetry sector $(N,Q,N_{\rm even},\mathcal{I})=(12, 134, 6, +1)$ whose dimension is 3477. (b) Distribution of the subsector size for various system sizes. (c) Site-number dependence of the number of frozen states. The blue solid line represents the fit to an exponential function. (d) Site-number dependence of the ratio between the dimension of the maximum subsector and dimension of the symmetry sector that belongs to the maximum subsector. The blue solid line represents the fit to an exponential function.}
\label{fig:hsf_figure_h1_r6}
\vspace{-0.75em}
\end{figure}%

\begin{figure}[t]
\centering
\includegraphics[width=8.5cm,clip]{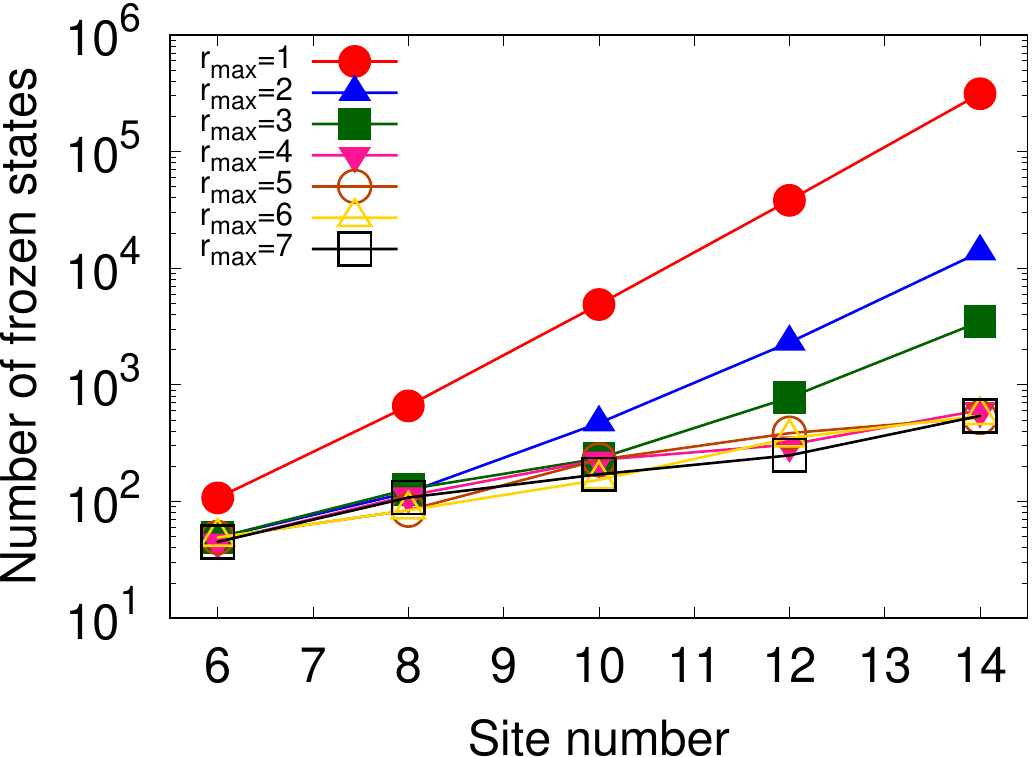}
\caption{System size dependence of the number of frozen states for $V=0.1J$ and $r_{\rm max}=1\text{-}7$.}
\label{fig:rmax_dependence_hsf_figure_h1_r7}
\vspace{-0.75em}
\end{figure}%

\section{ANALYTICAL ESTIMATION OF THE NUMBER OF FROZEN STATES}\label{sec:analytical_estimation_frozen_state}
Here, we estimate the number of frozen states for the soft-core boson Hamiltonian $\hat{H}=\hat{H}_0+\hat{H}_{\rm int}$ used in the main text. Because the present system has the space-inversion symmetry, the frozen state is defined by the state that satisfies $\hat{H}_{\rm int}\cket{\bar{\bm{n}},\pm}=E_0(\bar{\bm{n}})\cket{\bar{\bm{n}},\pm}$, where $E_0(\bar{\bm{n}})$ is a constant. We can categorize the frozen states into the following four classes. (i) A product state satisfies $\hat{H}_{\rm int}\cket{\bar{\bm{n}}}=0$. In this case, we can easily show $\hat{H}_{\rm int}\cket{\bar{\bm{n}},\pm}=\hat{H}_{\rm int}[\sqrt{q(\bar{\bm{n}})}/2](\cket{\bar{\bm{n}}}\pm\cket{I(\bar{\bm{n}})})=\pm \hat{H}_{\rm int}[\sqrt{q(\bar{\bm{n}})}/2]\hat{\mathcal{I}}\cket{\bar{\bm{n}}}=\pm[\sqrt{q(\bar{\bm{n}})}/2]\hat{\mathcal{I}}\hat{H}_{\rm int}\cket{\bar{\bm{n}}}=0,$ where we used the space-inversion symmetry. (ii) A product state satisfies $\hat{H}_{\rm int}\cket{\bar{\bm{n}}}=E_0(\bar{\bm{n}})\cket{I(\bar{\bm{n}})}$, where $E_0(\bar{\bm{n}})\not=0$. In this case, we can obtain $\hat{H}_{\rm int}\cket{\bar{\bm{n}},\pm}=[\sqrt{q(\bar{\bm{n}})}/2][E_0(\bar{\bm{n}})\cket{I(\bar{\bm{n}})}\pm E_0(\bar{\bm{n}})\cket{\bar{\bm{n}}}]=\pm E_0(\bar{\bm{n}})\cket{\bar{\bm{n}},\pm}$. (iii) The following relations hold: $\hat{H}_{\rm int}\cket{\bar{\bm{n}}}=E_0(\bar{\bm{n}},\bar{\bm{m}})\cket{\bar{\bm{m}}}$, where $\hat{\mathcal{I}}\cket{\bar{\bm{m}}}=\cket{\bar{\bm{m}}}$, $\cket{\bm{m}}\not=\cket{I(\bar{\bm{n}})}$, and $E_0(\bar{\bm{n}},\bar{\bm{m}})\not=0$. In this case, we can find the frozen state in $\mathcal{I}=-1$ sector: $\hat{H}_{\rm int}\cket{\bar{\bm{n}},-1}=[\sqrt{q(\bar{\bm{n}})}/2]\hat{H}_{\rm int}[\cket{\bar{\bm{n}}}-\cket{I(\bar{\bm{n}})}]=0$. (iv) Accidental case. For example, we find that $\cket{\bar{\bm{n}},-}=\cket{2,0,1,2,0,1,-}$ is a frozen state in $\mathcal{I}=-1$ sector. 

In the rest of this section, we analytically estimate the number of the frozen state for the cases (i) and (ii) because the estimation of the classes (iii) and (iv) are complicated and their number is relatively small compared to classes (i) and (ii) as shown below. We also note that the results presented below are valid for $M\ge 6$.

Before going into the details of the analysis, we rewrite the interaction Hamiltonian (\ref{eq:two-body_interaction_Hamiltonian}) in a convenient form for the analysis:
\begin{align}
\hat{H}_{\rm int}&=\frac{2V}{M}\sum_{l=0}^{M/2-1}(\hat{b}^{\dagger}_l\hat{b}^{\dagger}_{M-l-1}\hat{b}_{l+1}\hat{b}_{M-l}+\hat{b}^{\dagger}_{M-l}\hat{b}^{\dagger}_{l+1}\hat{b}_l\hat{b}_{M-l-1})\equiv \frac{2V}{M}\sum_{l=0}^{M/2-1}\hat{h}_l,\label{eq:rewrite_Hamiltonian_convenient_form}
\end{align}
where we changed the dummy indices and used $\hat{b}_l=\hat{b}_{l\pm M}$. From the periodicity of the annihilation and creations operators, the Hamiltonian has $M$ independent terms.  

Here, we count the number of frozen states for the class (i). In this case, $\hat{H}_{\rm int}\cket{\bm{n}}=0$ holds. $\bm{n}$ must satisfy the following  $M$ conditions:
\begin{align}
(\text{Condition 1}) &: n_1n_0=0,\quad n_2n_{M-1}=0,\ldots, n_{M/2-1}n_{M/2+1}=0,\quad n_{M/2}n_{M/2+1}=0,\label{eq:condition1_frozen_state}\\
(\text{Condition 2}) &: n_0n_{M-1}=0,\quad n_1n_{M-1}=0,\ldots, n_{M/2}n_{M/2+1}=0, \quad n_{M/2-1}n_{M/2}=0.\label{eq:condition2_frozen_state}
\end{align}
Equations (\ref{eq:condition1_frozen_state}) and (\ref{eq:condition2_frozen_state}) correspond to the conditions where all $\hat{h}_l $ in Eq.~(\ref{eq:rewrite_Hamiltonian_convenient_form}) vanish when we apply $\hat{H}_{\rm int}$ to $\cket{\bm{n}}$. The above conditions are complicated. To understand the conditions easily, we visualize the conditions in Fig.~\ref{fig:supple_frozen_state_class1}. From Fig.~\ref{fig:supple_frozen_state_class1}(b), the number of frozen states of class 1 corresponds to the number of ways for arranging $M$ nonnegative integers on a ring whose summation is $N$, and products of any adjacent numbers become zero. This number is given by
\begin{align}
D^{({\rm i})}(N,M)&=\sum_{p=1}^{M/2}\binom{M-1-p+1}{p}\binom{N-1}{p-1}+\left[1+\sum_{n_{M-1}=1}^{N-1}\sum_{p=1}^{M/2-1}\binom{M-3-p+1}{p}\binom{N-n_{M-1}-1}{p-1}\right],\label{eq:number_of_frozen_state_class1}
\end{align}
where $\binom{n}{m}\equiv n!/[(n-m)!m!]$ is a binomial constant ($n, m\ge 0$). The first term represents the case $n_{M-1}=0$ and the second term $n_{M-1}\not=0$. This result can be obtained as follows. 
First, we consider the case $n_{M-1}=0$. In this case, the number of ways of choosing $p$ numbers from $M-1$ numbers and satisfying the conditions 1 and 2 is given by $\binom{M-1-p+1}{p}$, where $p=1,2,\ldots,M/2$. The number of combinations that the summation of $p$ numbers becomes $N$ is given by $\binom{N-1}{p-1}$. The product of these numbers $\binom{M-1-p+1}{p}\binom{N-1}{p-1}$ represents the number of combinations that the $p$ positive number satisfies the conditions 1 and 2 and their summation becomes $N$. Therefore, we obtain the first term of Eq.~(\ref{eq:number_of_frozen_state_class1}). Next, we consider the case $n_{M-1}\not=0$. In this case $n_0$ and $n_2$ must be zero due to the conditions 1 and 2 [see Fig.~\ref{fig:supple_frozen_state_class1} (b)]. The number of ways of choosing $p$ numbers from the remaining $M-3$ number and satisfying the conditions 1 and 2 is given by $\binom{M-3-p+1}{p}$, where $p=1,2,\ldots, M/2-1$. For fixed $n_{M-1}\;(=1,2,\ldots, N)$, the number of combinations that the summation of $p$ numbers becomes $N-n_{M-1}$ is given by $\binom{N-n_{M-1}-1}{p-1}$. Taking the summation of $\binom{M-3-p+1}{p}\binom{N-n_{M-1}-1}{p-1}$ for all possible values $p$ and $n_{M-1}$, we obtain the second term of Eq.~(\ref{eq:number_of_frozen_state_class1}), where $+1$ represents the case $n_{M-1}=N$.

\begin{figure}[t]
\centering
\includegraphics[width=15cm,clip]{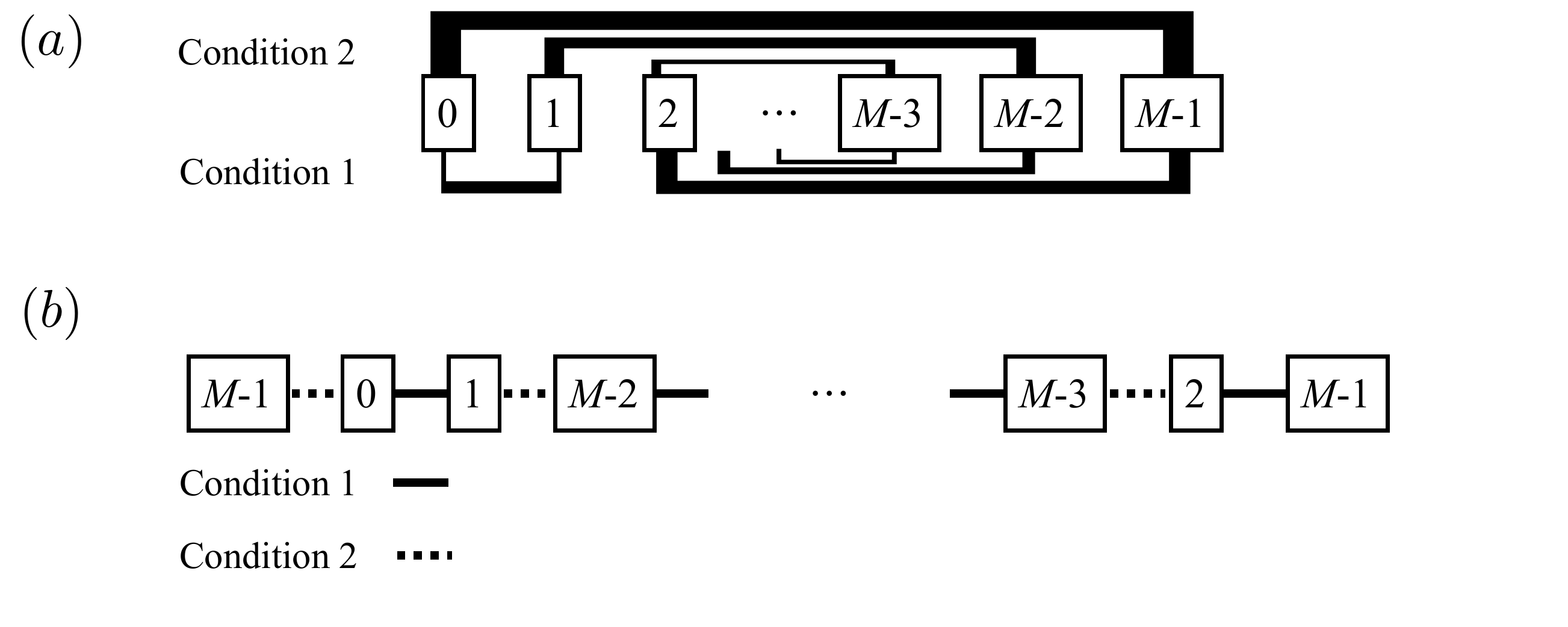}
\vspace{-1.0em}
\caption{Schematic figure for estimation of the class (i) frozen states. Each box represents $n_l$. (a) The black solid line represents the condition 1 or 2. The product of connected boxes is zero. (b) Rearrange the order of $n_l$. The black solid and dotted lines represent conditions 1 and 2, respectively.}
\label{fig:supple_frozen_state_class1}
\end{figure}%

Then, we consider the class (ii). In this case, $\hat{H}_{\rm int}\cket{\bm{n}}=E_0(\bm{n})\cket{I(\bm{n})}$. This means that, for the product state $\cket{\bm{n}}$, the interaction Hamiltonian couples only its space-inversion state $\cket{I(\bm{n})}$. The class (ii) is further categorized into five patterns. These patterns are characterized by the crystal momentum in which the matrix elements of the operator $\hat{h}_l$ are nonzero: $\hat{h}_{l=0}$ (pattern 1), $\hat{h}_{l=M/2-1}$ (pattern 2), $\hat{h}_{l=1}$ (pattern 3), $\hat{h}_{l=M/2-2}$ (pattern 4), and $\hat{h}_{l\not=0,1,M/2-2,M/2-1}$ (pattern 5). We will see these one by one. 

In pattern 1, the nonzero matrix element is given by $\hat{h}_{l=0}$. In this case, we obtain the following conditions:
\begin{align}
&n_0>0,\quad (n_1,n_{M-1})=(0,1)\text{ or }(1,0),\quad n_2=n_{M-2}=0,\label{eq:condition1_class2_pattern1}\\
&\text{For other }l,\text{ Eqs.~(\ref{eq:condition1_frozen_state}) and (\ref{eq:condition2_frozen_state}) are satisfied and }n_l=n_{M-l}.\label{eq:condition2_class2_pattern1}
\end{align}
See also Fig.~\ref{fig:supple_frozen_state_class2_pattern1}. From these conditions, we need to consider the combination such that $2(n_3+n_4+\cdots+n_{M/2-3})=N-n_0-n_{M/2}-1\equiv N'$. Because $N'$ must be an even number, we have four cases: $(N,n_0,n_{M/2})=({\rm even}, {\rm even}, {\rm odd}),\;({\rm even}, {\rm odd}, {\rm even}),\;({\rm odd}, {\rm even}, {\rm even}),$ and $({\rm odd}, {\rm odd}, {\rm odd})$. The same approach as in the class (i) yields the following results:
\begin{align}
D^{({\rm ii})}_1(N,M)&\equiv 
\begin{cases}
\vspace{0.5em}D^{({\rm ii})}_{1,({\rm e,e,o})}(N,M)+D^{({\rm ii})}_{1,({\rm e,o,e})}(N,M),\quad (\text{even }N),\\
D^{({\rm ii})}_{1,({\rm o,e,e})}(N,M)+D^{({\rm ii})}_{1,({\rm o,o,o})}(N,M),\quad (\text{odd }N),
\end{cases}
\label{eq:class2_pattern1}
\end{align}
\begin{align}
D^{({\rm ii})}_{1,({\rm e,e,o})}(N,M)&=2\sum_{n_0=2,4,\ldots}^{N-2}\left[1+\sum_{n_{M/2}=1,3,\ldots}^{N-3-n_0}\sum_{p=1}^{1+\lfloor (M-8)/4\rfloor}\binom{(M-8)/2-p+1}{p}\binom{(N-1-n_0-n_{M/2})/2-1}{p-1}\right],\label{eq:class2_pattern1_eeo}\\
D^{({\rm ii})}_{1,({\rm e,o,e})}(N,M)&=2+2\sum_{n_0=1,3,\ldots}^{N-3}\sum_{p=1}^{1+\lfloor (M-6)/4\rfloor}\binom{(M-6)/2-p+1}{p}\binom{(N-1-n_0)/2-1}{p-1}\notag \\
&\quad +2\sum_{n_0=1,3,\ldots}^{N-3}\left[1+\sum_{n_{M/2}=2,4,\ldots}^{N-3-n_0}\sum_{p=1}^{1+\lfloor(M-8)/4\rfloor}\binom{(M-8)/2-p+1}{p}\binom{(N-1-n_0-n_{M/2})/2-1}{p-1}\right],\label{eq:class2_pattern1_eoe}\\
D^{({\rm ii})}_{1,({\rm o,e,e})}(N,M)&=2+2\sum_{n_0=2,4,\ldots}^{N-3}\sum_{p=1}^{1+\lfloor(M-6)/4\rfloor}\binom{(M-6)/2-p+1}{p}\binom{(N-1-n_0)/2-1}{p-1}\notag \\
&\quad +2\sum_{n_0=2,4,\ldots}^{N-3}\left[1+\sum_{n_{M/2}=2,4,\ldots}^{N-3-n_0}\sum_{p=1}^{1+\lfloor(M-8)/4\rfloor}\binom{(M-8)/2-p+1}{p}\binom{(N-1-n_0-n_{M/2})/2-1}{p-1}\right],\label{eq:class2_pattern1_oee}\\
D^{({\rm ii})}_{1,({\rm o,o,o})}(N,M)&=2\sum_{n_0=1,3,\ldots}^{N-2}\left[1+\sum_{n_{M/2}=1,3,\ldots}^{N-3-n_0}\sum_{p=1}^{1+\lfloor(M-8)/4\rfloor}\binom{(M-8)/2-p+1}{p}\binom{(N-1-n_0-n_{M/2})/2-1}{p-1}\right],\label{eq:class2_pattern1_ooo}
\end{align}
where $\lfloor \cdot\rfloor$ represents the floor function.

In pattern 2, the nonzero matrix element is given by $\hat{h}_{l=M/2-1}$. In this case, we obtain the following conditions:
\begin{align}
&n_{M/2}>0,\quad (n_{M/2+1},n_{M/2-1})=(0,1)\text{ or }(1,0),\quad n_{M/2-2}=n_{M/2+2}=0,\label{eq:condition1_class2_pattern2}\\
&\text{For other }l,\text{ Eqs.~(\ref{eq:condition1_frozen_state}) and (\ref{eq:condition2_frozen_state}) are satisfied and }n_l=n_{M-l}.\label{eq:condition2_class2_pattern2}
\end{align}
We can find that the number of frozen state for this pattern is same as that of pattern 1. Therefore, we obtain
\begin{align}
D_2^{({\rm ii})}(N,M)=D_1^{({\rm ii})}(N,M).\label{eq:class2_pattern2}
\end{align}

\begin{figure}[t]
\centering
\includegraphics[width=15cm,clip]{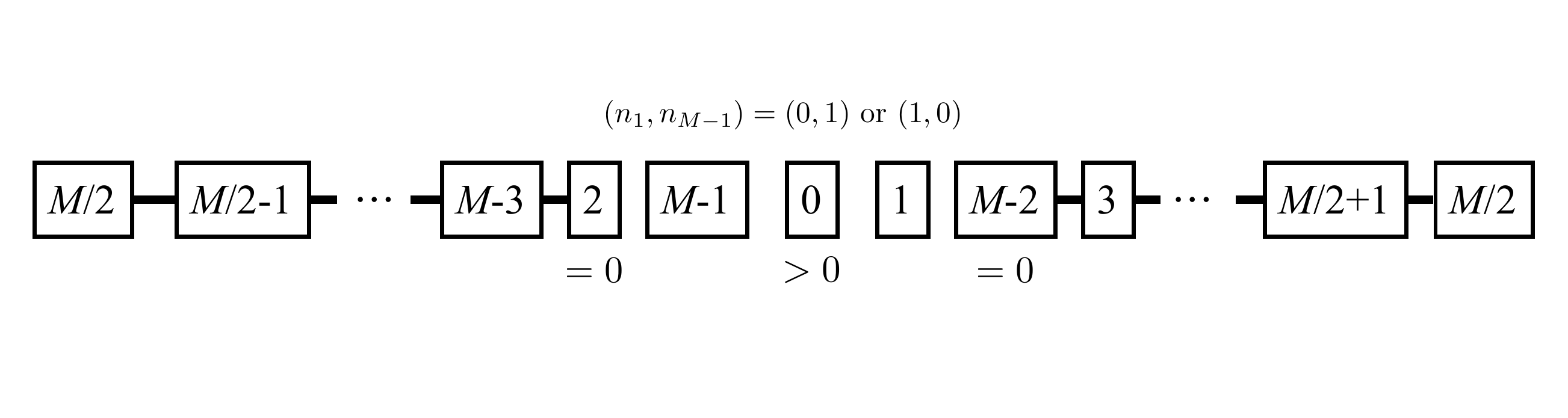}
\vspace{-4.0em}
\caption{Schematic figure for estimation of the class (ii) pattern 1 frozen states. Each box represents $n_l$.}
\label{fig:supple_frozen_state_class2_pattern1}
\end{figure}%

In pattern 3, the nonzero matrix element is given by $\hat{h}_{l=1}$. In this case, we obtain the following conditions:
\begin{align}
&n_0=n_3=n_{M-3}=0,\text{ and }\{[(n_1,n_{M-1})=(0,1)\text{ and }n_2=n_{M-2}+1],\text{or }[(n_1,n_{M-1})=(1,0)\text{ and }n_2=n_{M-2}-1],\notag\\
&\text{or }[(n_2,n_{M-2})=(1,0)\text{ and }n_1=n_{M-1}-1],\text{or }[(n_2,n_{M-2})=(0,1)\text{ and }n_1=n_{M-1}+1]\},\notag\\
&\text{and for other }l,\text{ Eqs.~(\ref{eq:condition1_frozen_state}) and (\ref{eq:condition2_frozen_state}) are satisfied and }n_l=n_{M-l}.\label{eq:condition_class2_pattern3}
\end{align}
See also Fig.~\ref{fig:supple_frozen_state_class2_pattern3}. We note that the double counting happens when $(n_1,n_{M-1},n_2,n_{M-2})=(0,1,1,0),\text{ or }(1,0,0,1)$. Using the condition that $(N,n_{M/2})=({\rm even}, {\rm even}),\text{ or }({\rm odd},{\rm odd})$, we obtain
\begin{align} 
D_3^{({\rm ii})}(N,M)&\equiv 
\begin{cases}
\vspace{0.5em}4,\quad M=6,\\
\vspace{0.5em}D^{({\rm ii)}}_{3,{\rm even}}(N,M),\quad (\text{even }N \text{ and } M>6),\\
D^{({\rm ii)}}_{3,{\rm odd}}(N,M),\quad (\text{odd }N \text{ and } M>6),\\
\end{cases}
\label{eq:class2_pattern3}
\end{align}
\begin{align}
&D^{({\rm ii)}}_{3,{\rm even}}(N,M)\notag \\
&=2\sum_{n_{M-2}=1}^{N/2-2}\left[1+\sum_{n_{M/2}=2,4,\ldots}^{N-2n_{M-2}-4}\sum_{p=1}^{1+\lfloor (M-10)/4\rfloor}\binom{(M-10)/2-p+1}{p}\binom{(N-n_{M/2}-2n_{M-2}-2)/2-1}{p-1}\right]\notag \\
&\quad +2+2\sum_{n_{M-2}=1}^{N/2-2}\sum_{p=1}^{1+\lfloor(M-8)/4\rfloor}\binom{(M-8)/2-p+1}{p}\binom{(N-2n_{M-2}-2)/2-1}{p-1}\notag \\
&\quad +1+\sum_{n_{M/2}=2,4,\ldots}^{N-4}\sum_{p=1}^{1+\lfloor(M-10)/4\rfloor}\binom{(M-10)/2-p+1}{p}\binom{(N-n_{M/2}-2)/2-1}{p-1}\notag \\
&\quad+\sum_{p=1}^{1+\lfloor(M-8 )/4\rfloor}\binom{(M-8)/2-p+1}{p}\binom{(N-2)/2-1}{p-1}\notag \\
&\quad +2\sum_{n_{M-2}=2}^{N/2-1}\left[1+\sum_{n_{M/2}=2,4,\ldots}^{N-2n_{M-2}-2}\sum_{p=1}^{1+\lfloor(M-10)/4\rfloor}\binom{(M-10)/2-p+1}{p}\binom{(N-n_{M/2}-2n_{M-2})/2-1}{p-1}\right]\notag \\
&\quad +2+2\sum_{n_{M-2}=2}^{N/2-1}\sum_{p=1}^{1+\lfloor(M-8)/4\rfloor}\binom{(M-8)/2-p+1}{p}\binom{(N-2n_{M-2})/2-1}{p-1}\notag \\
&\quad +1+\sum_{n_{M/2}=2,4,\ldots}^{N/2-1}\sum_{p=1}^{1+\lfloor(M-10)/4\rfloor}\binom{(M-10)/2-p+1}{p}\binom{(N-n_{M/2}-2)/2-1}{p-1}\notag \\
&\quad +\sum_{p=1}^{1+\lfloor(M-8)/4\rfloor}\binom{(M-8)/2-p+1}{p}\binom{(N-2)/2-1}{p-1},\label{eq:class2_pattern3_evenN}
\end{align}
\begin{figure}[t]
\centering
\includegraphics[width=15cm,clip]{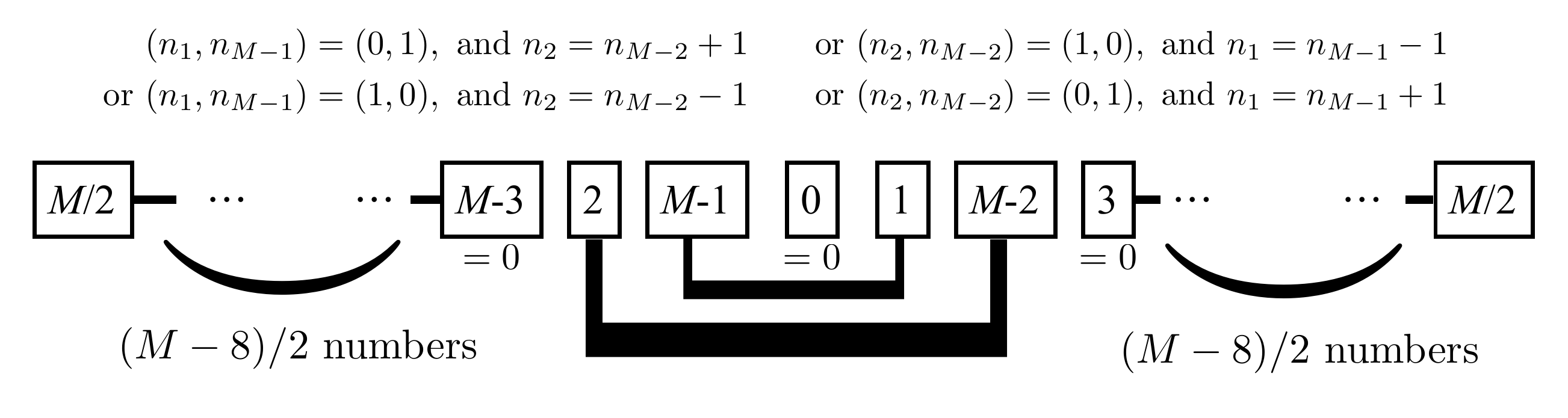}
\caption{Schematic figure for estimation of the class (ii) pattern 3 frozen states. Each box represents $n_l$.}
\label{fig:supple_frozen_state_class2_pattern3}
\end{figure}%
\begin{align}
&D^{({\rm ii)}}_{3,{\rm odd}}(N,M)\notag \\
&=2+2\sum_{n_{M-2}=1}^{(N-5)/2}\left[1+\sum_{n_{M/2}=1,3,\ldots}^{N-2n_{M-2}-4}\sum_{p=1}^{1+\lfloor(M-10)/4\rfloor}\binom{(M-10)/2-p+1}{p}\binom{(N-n_{M/2}-2n_{M-2}-2)/2-1}{p-1}\right]\notag \\
&+1+\sum_{n_{M/2}=1,3,\ldots}^{N-4}\sum_{p=1}^{1+\lfloor(M-10)/4\rfloor}\binom{(M-10)/2-p+1}{p}\binom{(N-n_{M/2})/2-1}{p-1}\notag \\
&+2+2\sum_{n_{M-2}=2}^{(N-3)/2}\left[1+\sum_{n_{M/2}=1,3,\ldots}^{N-2n_{M-2}-2}\sum_{p=1}^{1+\lfloor(M-10)/4\rfloor}\binom{(M-10)/2-p+1}{p}\binom{(N-n_{M/2}-2n_{M-2})/2-1}{p-1}\right]\notag \\
&+1+\sum_{n_{M/2}=1,3,\ldots}^{N-4}\sum_{p=1}^{1+\lfloor(M-10)/4\rfloor}\binom{(M-10)/2-p+1}{p}\binom{(N-n_{M/2}-2)/2-1}{p-1}.\label{eq:class2_pattern3_oddN}
\end{align}

In pattern 4, the nonzero matrix element is given by $\hat{h}_{l=M/2-2}$. In this case, we can find that the number of frozen states is same as pattern 3. Therefore, we obtain
\begin{align}
D^{({\rm ii})}_4(N,M)&=D^{({\rm ii})}_3(N,M).\label{eq:class2_pattern4}
\end{align}

Finally, we consider pattern 5. In this case, the nonzero matrix element is given by $\hat{h}_{l\not=0,1,M/2-2,M/2-1}$. The conditions are given by
\begin{align}
n_{M-l+1}&=n_{l+2}=n_{l-1}=n_{M-l-2}=0, \text{ and}\left\{[(n_l,n_{M-l})=(1,0)\text{ and }n_{l+1}=n_{M-l-1}+1], \right.\notag \\
&\left.\text{ or }[(n_{l+1},n_{M-l-1})=0\text{ and }n_l=n_{M-l}-1],\text{ or }[(n_l, n_{M-l-1})=(1,0)\text{ and }n_{l+1}=n_{M-l-1}-1],\right.\notag \\
&\left. \text{ or }[(n_{l+1},n_{M-l-1})=(0,1)\text{ and }n_l=n_{M-l}+1]\right\},\notag \\
&\text{ and for other }l,\text{ Eqs.~(\ref{eq:condition1_frozen_state}) and (\ref{eq:condition2_frozen_state}) are satisfied and }n_l=n_{M-l}.\label{eq:condition_class2_pattern5}
\end{align}
See also Fig.~\ref{fig:supple_frozen_state_class2_pattern5}. We note that the double counting happens when $(n_l,n_{M-l},n_{l+1},n_{M-l-1})=(0,1,1,0),\text{ or }(1,0,0,1)$ as in pattern 3, and this pattern appears when $M\ge 10$. To obtain the expression of the number of combination, we use the following function:
\begin{align}
D_0(N_0,M_0)&\equiv 
\begin{cases}
\vspace{0.2em}1,\quad M_0\le 0 \text{ or } N_0=0,\\
\vspace{0.2em}D_0^{\rm e}(N_0,M_0),\quad \text{even }N_0,\\
D_0^{\rm o}(N_0,M_0),\quad \text{odd }N_0,
\end{cases}
\label{eq:definition_of_D0}\\
D_0^{\rm e}(N_0,M_0)&\equiv 1+\sum_{m_0=2,4,\ldots}^{N_0-2}\sum_{p=1}^{1+\lfloor M_0/2-1\rfloor}\binom{M_0-2+p}{p}\binom{(N_0-m_0)/2-1}{p-1}+\sum_{p=1}^{1+\lfloor M_0/2\rfloor}\binom{M_0-1+p}{p}\binom{N_0/2-1}{p-1},\label{eq:definition_D0_even_N0}\\
D_0^{\rm o}(N_0,M_0)&\equiv 1+\sum_{m_0=1,3,\ldots}^{N_0-2}\sum_{p=1}^{1+\lfloor M_0/2-1\rfloor}\binom{M_0-2+p}{p}\binom{(N_0-m_0)/2-1}{p-1}.\label{eq:definition_D0_odd_N0}
\end{align}
$D_0^{\rm e}(N_0,M_0)$ and $D_0^{\rm o}(N_0,M_0)$ represent the number of combinations that satisfies $N_0=m_0+2(m_1+m_2+\ldots+m_{M_0})$, where $N_0, M_0, m_1,\cdots, m_{M_0}$ are nonnegative integers. For fixed $l$, the number of the combinations is given by
\begin{align}
D(l,N,M)&=
\begin{cases}
\vspace{0.2em}0,\quad \text{for } M<10,\\
\vspace{0.2em}D^{\rm e}(l,N,M),\quad \text{even }N,\\
D^{\rm o}(l,N,M),\quad \text{odd }N,\\
\end{cases}
\label{eq:number_of_combination_for_fixed_l}\\
D^{\rm e}(l,N,M)&\equiv 2\sum_{n_{M-l-1}=1}^{N/2-1}\sum_{k=0}^{N-2n_{M-l-1}-2}D_0(k,l-2)D_0(N-2n_{M-l-1}-2-k,M/2-l-3)\notag \\
&\quad +2\sum_{n_{M-l-1}=2}^{N/2}\sum_{k=0}^{n-2n_{M-l-1}}D_0(k,l-2)D_0(N-2n_{M-l-1}-k,M/2-l-3)\notag \\
&\quad +\sum_{k=0}^{N-2}D_0(k,l-2)D_0(N-2-k,M/2-l-3)+\sum_{k=0}^{N-2}D_0(k,l-2)D_0(N-2-k,M/2-l-3),\label{eq:definition_of_De_fixed_l}
\end{align}
\begin{align}
D^{\rm o}(l,N,M)&\equiv 2\sum_{n_{M-l-1}=1}^{(N-1)/2-1}\sum_{k=0}^{N-2n_{M-l-1}-2}D_0(k,l-2)D_0(N-2n_{M-l-1}-2-k,M/2-l-3)\notag \\
&\quad +2\sum_{n_{M-l-1}=2}^{(N-1)/2}\sum_{k=0}^{n-2n_{M-l-1}}D_0(k,l-2)D_0(N-2n_{M-l-1}-k,M/2-l-3)\notag \\
&\quad +\sum_{k=0}^{N-2}D_0(k,l-2)D_0(N-2-k,M/2-l-3)+\sum_{k=0}^{N-2}D_0(k,l-2)D_0(N-2-k,M/2-l-3).\label{eq:definition_of_Do_fixed_l}
\end{align}
Therefore, we obtain 
\begin{align}
D_5^{({\rm ii})}(N,M)&=\sum_{l=2}^{M/2-3}D(l,N,M).\label{eq:class2_pattern5}
\end{align}

\begin{figure}[t]
\centering
\includegraphics[width=15cm,clip]{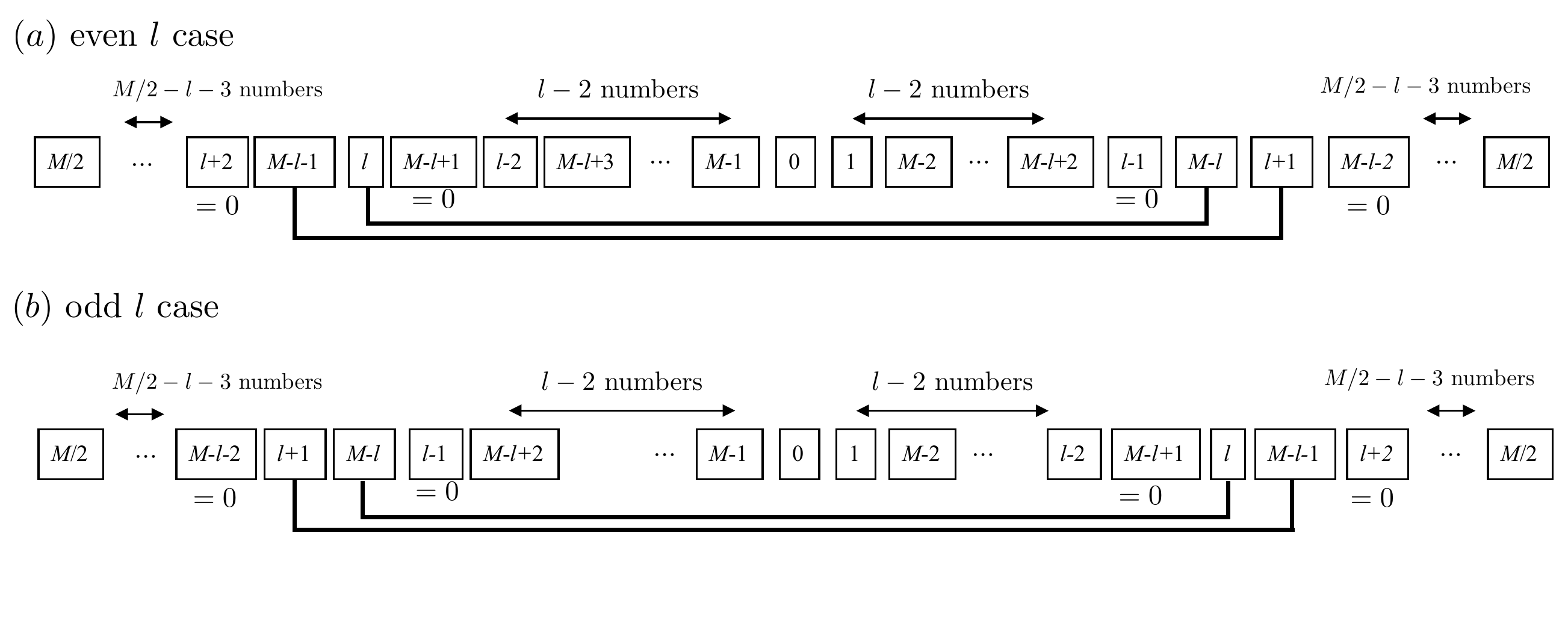}
\vspace{-2.0em}
\caption{Schematic figure for estimation of the class (ii) pattern 5 frozen states. Each box represents $n_l$.}
\label{fig:supple_frozen_state_class2_pattern5}
\end{figure}%

From the above results, we obtained the number of frozen states of class (i) and (ii):
\begin{align}
D^{{\rm (i),(ii)}}(N,M)\equiv D^{(i)}(N,M)+\sum_{j=1}^5D^{{\rm (ii)}}_j(N,M).\label{eq:frozen_state_class_i_and_ii}
\end{align}
Figure \ref{fig:ratio_numerical_analytical_results} shows the ratio between the analytical results (\ref{eq:frozen_state_class_i_and_ii}) and numerical results for $M\le 14$. We find that the ratio $D^{{\rm (i),(ii)}}/D^{\rm numerical}$ approaches 1 as the system size increases. For $M=14$, the ratio is roughly over 99\%. From this result, analytical result (\ref{eq:frozen_state_class_i_and_ii}) approximates the correct result well. Therefore, we fit the analytical results to the exponential function. The fitting region is $14<M\le 100$. The results are shown in Fig.~\ref{fig:Compare_numerical_analytical_results}. Because the analytical result (\ref{eq:frozen_state_class_i_and_ii}) can be regarded as the lower bound of the number of frozen states, this result supports that the number of frozen states increases exponentially with the system size, which suggests the strong HSF occurs in this system.

\begin{figure}[t]
\centering
\includegraphics[width=10cm,clip]{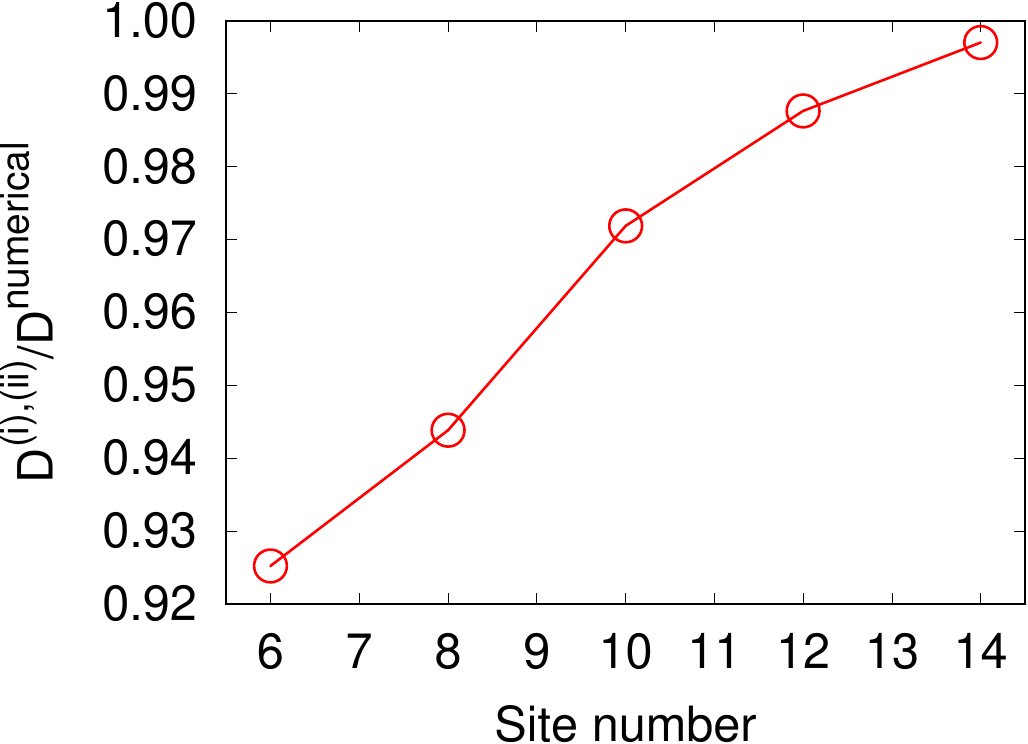}
\caption{The ratio between the number of frozen states estimated by numerical and analytical ways.}
\label{fig:ratio_numerical_analytical_results}
\end{figure}%

\begin{figure}[t]
\centering
\includegraphics[width=10cm,clip]{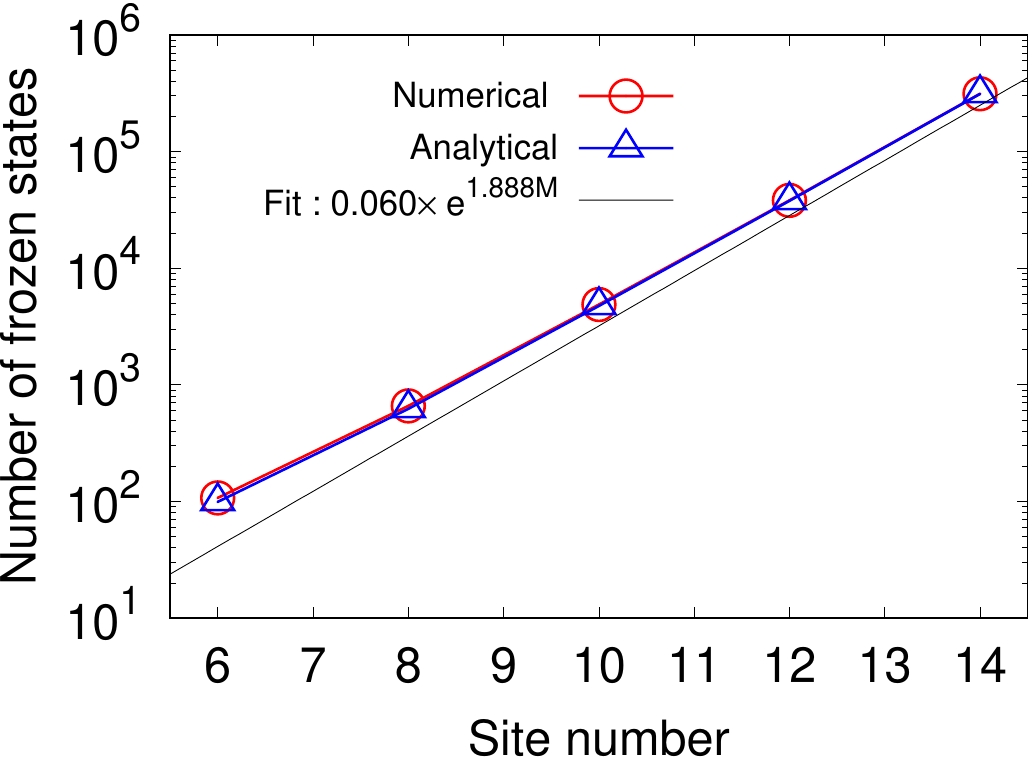}
\caption{Site number dependence of the number of frozen state for numerical (red circle) and analytical (blue triangle) results. The black solid line represents the fitting result. }
\label{fig:Compare_numerical_analytical_results}
\end{figure}%

\section{ANALYTICAL ESTIMATION OF THE NUMBER OF FINITE-CURRENT STATES}\label{sec:analytical_estimation_finite-current}

Here, we estimate the number of the finite-current states. The Hamiltonian is the same as that in Appendix~\ref{sec:analytical_estimation_frozen_state}. There are two kinds of finite-current states. One is the stationary-current states and the other one is oscillating current states.

As discussed in the main text, the stationary-current states appear when $E_+=E_-$, where $E_{\pm}$ is the eigenenergy of the frozen states $\cket{\bar{\bm{n}}, \pm}$. This situation occurs when $\hat{H}_{\rm int}\cket{\bar{\bm{n}}}=\hat{H}_{\rm int}\cket{I(\bar{\bm{n}})}=0$ and $\hat{\mathcal{I}}\cket{\bar{\bm{n}}}\not\propto\cket{\bar{\bm{n}}}$. From these conditions, we can write down the expression of the number of the finite-current states $D_{\rm SC}(N,M)$:
\begin{align}
D_{\rm SC}(N,M)&=D^{({\rm i})}(N,M)-D_{\rm sym}(N,M),
\end{align}
where $D_{\rm sym}(N,M)$ is the number of frozen states that satisfy $\hat{\mathcal{I}}\cket{\bm{n}}=\cket{\bm{n}}$ and $\hat{H}_{\rm int}\cket{\bm{n}}=0$. We call these states symmetric frozen states. The symmetric frozen state must satisfy the conditions (\ref{eq:condition1_frozen_state}), (\ref{eq:condition2_frozen_state}), and $n_l=n_{M-l}$ for $l=1,2,\ldots, M/2-1$. We can obtain the expression of $D_{\rm sym}(N,M)$ as follows:
\begin{align}
D_{\rm sym}(N,M)&=
\begin{cases}
D_{\rm sym,e}(N,M),\quad \text{even }N,\\
D_{\rm sym,o}(N,M),\quad \text{odd }N,
\end{cases}
\label{eq:number_of_symmetric_frozen_state}\\
D_{\rm sym,e}(N,M)&=N+1+\sum_{p=1}^{1+\lfloor(M/2-2)/2\rfloor}\binom{M/2-p}{p}\binom{N/2-1}{p-1}\notag \\
&\quad +2\sum_{k=2,4,\ldots,}^{N-2}\sum_{p=1}^{1+\lfloor(M/2-3)/2\rfloor}\binom{M/2-p-1}{p}\binom{(N-k)/2-1}{p-1}\notag \\
&\quad +\sum_{k=2,4,\ldots,}^{N-2}\sum_{p=1}^{1+\lfloor(M/2-4)/2\rfloor}(k-1)\binom{M/2-p-2}{p}\binom{(N-k)/2-1}{p-1},\label{eq:total_number_of_symmetric_frozen_state_for_even_N}\\
D_{\rm sym,o}(N,M)&=N+1+2\sum_{k=1,3,\ldots}^{N-2}\sum_{p=1}^{1+\lfloor(M/2-3)/2\rfloor}\binom{M/2-p-1}{p}\binom{(N-k)/2-1}{p-1}\notag \\
&\quad +\sum_{k=1,3,\ldots}^{N-2}\sum_{p=1}^{1+\lfloor(M/2-4)/2\rfloor}(k-1)\binom{M/2-p-2}{p}\binom{(N-k)/2-1}{p-1}.\label{eq:total_number_of_symmetric_frozen_state_for_odd_N}
\end{align}

We can also estimate the number of frozen states that satisfy $E_{+}\not=E_-$. We denote this number as $D_{\rm OSC}(N,M)$. These states yield the periodically oscillating current states as discussed in the main text. This situation occurs when $\hat{H}_{\rm int}\cket{\bar{\bm{n}}}=E_0(\bar{\bm{n}})\cket{I(\bar{\bm{n}})}$ and $E_0(\bar{\bm{n}})\not=0$. From this condition, we can show $E_+\not=E_-$ as follows:
\begin{align}
\hat{H}\cket{\bar{\bm{n}},\pm}&=(\hat{H}_0+\hat{H}_{\rm int})\frac{1}{\sqrt{2}}[\cket{\bar{\bm{n}}}\pm\cket{I(\bar{\bm{n}})}]\notag \\
&=E_{\rm hop}(\bar{\bm{n}})\frac{1}{\sqrt{2}}[\cket{\bar{\bm{n}}}\pm\cket{I(\bar{\bm{n}})}]+E_0(\bar{\bm{n}})\frac{1}{\sqrt{2}}[\cket{I(\bar{\bm{n}})}\pm\cket{\bar{\bm{n}}}]\notag \\
&=[E_{\rm hop}(\bar{\bm{n}})\pm E_0(\bar{\bm{n}})]\cket{\bm{n},\pm},\label{eq:proof_E_+_is_not_equal_to_E_-}
\end{align}
where we used $\hat{H}_0\cket{\bar{\bm{n}}}=E_{\rm hop}(\bar{\bm{n}})\cket{\bar{\bm{n}}}$ and $\hat{H}_0\cket{I(\bar{\bm{n}})}=E_{\rm hop}(\bar{\bm{n}})\cket{I(\bar{\bm{n}})}$. The number of these states is equal to the number of the frozen states in class (ii). The expression of $D_{\rm OSC}(N,M)$ is given by
\begin{align}
D_{\rm OSC}(N,M)=\sum_{j=1}^5D_j^{({\rm ii})}(N,M).\label{eq:number_of_oscillating_frozen_state}
\end{align}
 
\begin{figure}[t]
\centering
\includegraphics[width=10cm,clip]{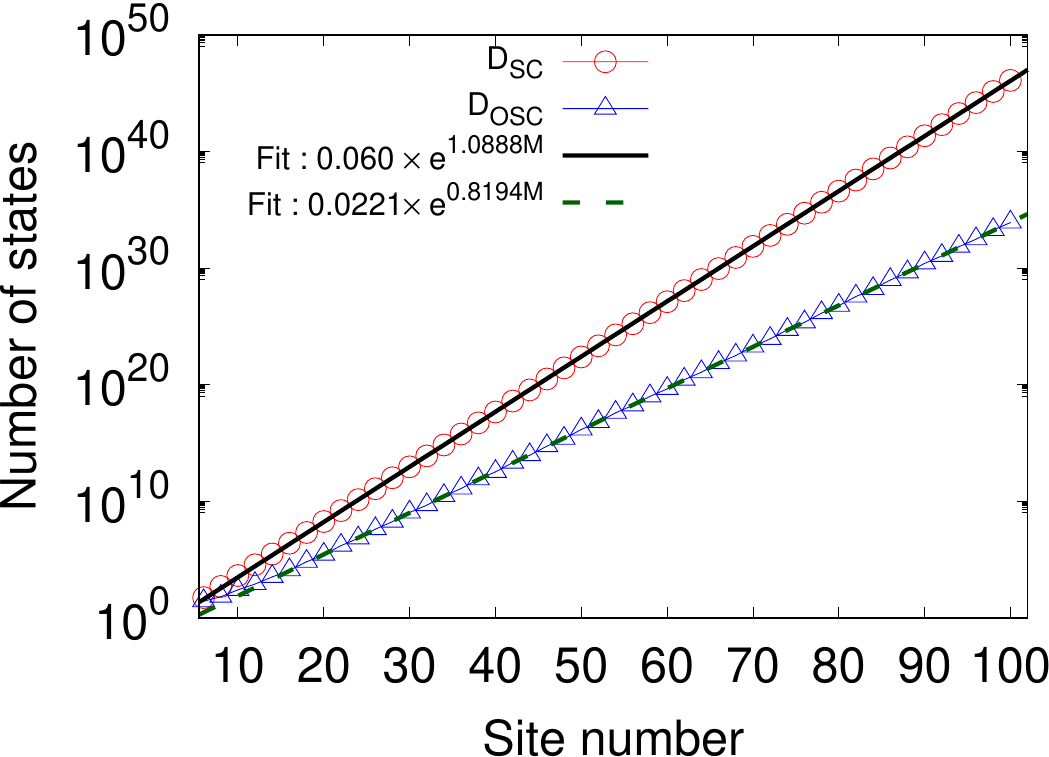}
\caption{Site number dependence of $D_{\rm SC}(N,M)$ (red circle) and $D_{\rm OSC}(N,M)$ (blue triangle) for $N=M$ cases. The black solid and green dashed lines represent the fitting results.}
\label{fig:number_of_pc_states}
\end{figure}%

We plot $D_{\rm SC}(N,M)$ and $D_{\rm OSC}(N,M)$ for $N=M$ cases in Fig.~\ref{fig:number_of_pc_states}. We also perform the fitting to these results with an exponential function. The fitting region is $14<M\le 100$. We can find that these quantities exponentially increase with the site number.

\section{DETAILS OF THE FITTING OF THE DECAY RATE}\label{sec:details_of_fitting}
Here, we explain how to extract the decay rate from the fitting to the fidelity. To perform the fitting to the fidelity properly, we need to specify the fitting region. For the short time regime, the fidelity can be written as
\begin{align}
|\bracket{\psi(0)}{\psi(t)}|^2\simeq 1-\left[\bra{\psi(0)}\hat{H}^2\cket{\psi(0)}-\bra{\psi(0)}\hat{H}\cket{\psi(0)}^2\right]\frac{t^2}{\hbar^2}.\label{eq:fidelity_short_time_region}
\end{align}
This means that the fidelity for the short time region is determined by the variance of the energy. We can also show that the long-time average of the fidelity converges to the inverse partition ratio \cite{Torres-Herrera2014}. In the long time regime, the fidelity fluctuates around this value. To extract the exponential decay, an appropriate choice of the fitting region is required.

Figure~\ref{fig:supple_fitting} shows an example of the fitting region. We summarize the fitting region $[t_1, t_2]$ as a function of the disorder strength in Table~\ref{tab:fitting_region}. We used these values to obtain the data shown in Fig.~\ref{fig:gamma_0_10}.

\begin{figure}[t]
\centering
\includegraphics[width=10cm,clip]{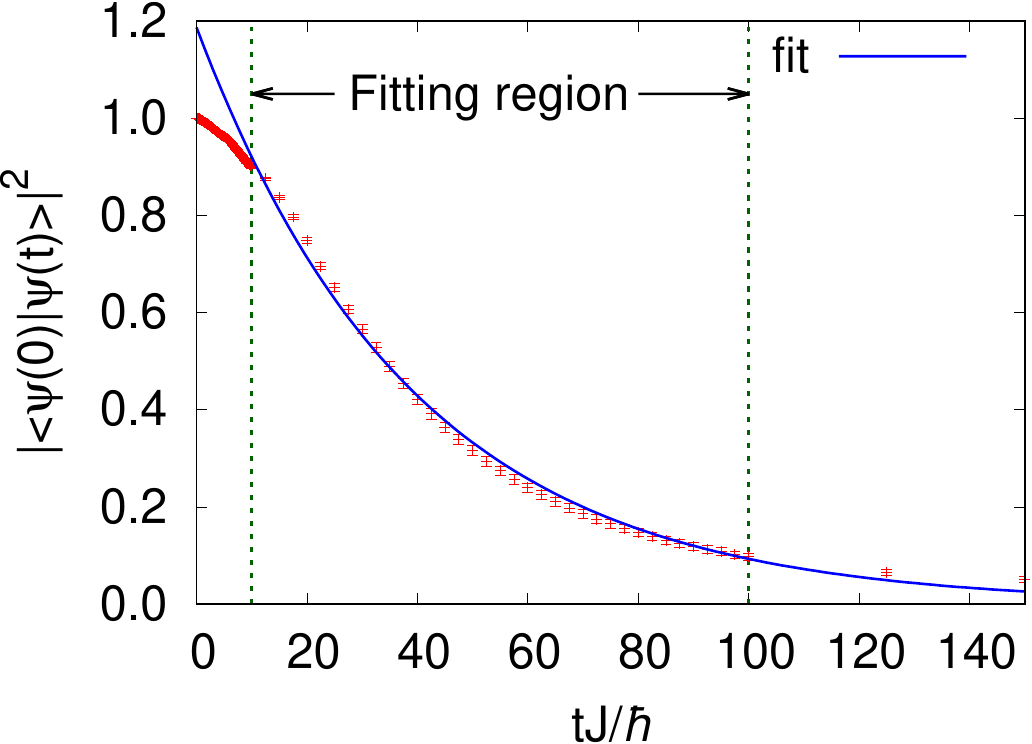}
\caption{Averaged fidelity for $M=N=10$, $V=0.1J$, and $W=0.1J$. In this case, the fitting region is $10\hbar/J\le t\le 100\hbar/J$. The solid blue line represents the fit to a function $Ae^{-\Gamma t}$.}
\label{fig:supple_fitting}
\vspace{-0.75em}
\end{figure}%

\begin{table}[H]
\centering
\caption{List of the values of the fitting region.}
\begin{tabular}{|ccc|}\hline
$W/J$ & $t_1J/\hbar$ & $t_2J/\hbar$   \\ \hline
0.01& 100 & 1000\\ 
0.03& 100 & 300 \\ 
0.1& 10 & 100 \\ 
0.3& 5 & 30     \\ 
1.0& 0.5 & 1.5 \\ 
3.0&  0.2 &1.0 \\ 
10.0& 0.1 & 0.5\\ \hline
\end{tabular}
\label{tab:fitting_region}
\end{table}

\end{widetext}

\end{document}